\documentclass[%
reprint,
superscriptaddress,
amsmath,amssymb,
aps,
]{revtex4-2}

\usepackage{graphicx}
\usepackage{dcolumn}
\usepackage{bm}
\usepackage[%
pdftoolbar= true, %
pdfmenubar	=true, 
pdftitle	={Room-temperature ladder-type optical memory compatible with single photons from InGaAs quantum dots}, 
pdfsubject	={Text}, 
pdfauthor	={Maass, Ewald}, 
pdfkeywords	={Text}, 
pdfcreator	={Text}, 
pdfproducer	={Text}, 
colorlinks = true, 
linkcolor = black, 
citecolor = black, 
urlcolor = black, 
    ]{hyperref} 

\usepackage{comment}
\usepackage{xcolor} 
\usepackage{lipsum}  

\usepackage[T1]{fontenc} 
\usepackage{float} 
\graphicspath{{./Figures/}} 

\begin{document}

\preprint{APS/123-QED} 

\title{Room-temperature ladder-type optical memory compatible with single photons from 
semiconductor quantum dots}%

\author{Benjamin Maa\ss}
 \email{benjamin.maass@dlr.de}
 \affiliation{Institute of Optical Sensor Systems, German Aerospace Center (DLR), Rutherfordstra{\ss}e 2, 12489 Berlin, Germany.}%
 \affiliation{Institute of Optics and Atomic Physics (IOAP), Technische Universit\"at Berlin,  Hardenbergstra{\ss}e 36, 10623 Berlin, Germany.}
\author{Norman Vincenz Ewald}%
 \affiliation{Institute of Optical Sensor Systems, German Aerospace Center (DLR), Rutherfordstra{\ss}e 2, 12489 Berlin, Germany.}%
 \affiliation{Institute of Solid State Physics (IFKP), Technische Universit\"at Berlin, Hardenbergstra{\ss}e 36, 10623 Berlin, Germany.}

 \author{Avijit Barua}
 \affiliation{Institute of Solid State Physics (IFKP), Technische Universit\"at Berlin, Hardenbergstra{\ss}e 36, 10623 Berlin, Germany.}
  \author{Stephan Reitzenstein}
 \affiliation{Institute of Solid State Physics (IFKP), Technische Universit\"at Berlin,  Hardenbergstra{\ss}e 36, 10623 Berlin, Germany.}
 \author{Janik Wolters}
 \affiliation{Institute of Optical Sensor Systems, German Aerospace Center (DLR), Rutherfordstra{\ss}e 2, 12489 Berlin, Germany.}%
 \affiliation{Institute of Optics and Atomic Physics (IOAP), Technische Universit\"at Berlin, Hardenbergstra{\ss}e 36, 10623 Berlin, Germany.}
\homepage{berlinquantum.net}

\date{\today}

\begin{abstract}
On-demand storage and retrieval of quantum information in coherent light-matter interfaces is a key requirement for future quantum networking and quantum communication applications. Alkali vapor memories offer scalable and robust high-bandwidth storage at high repetition rates, which makes them a natural fit to interface with solid-state single-photon sources. 
Here, we experimentally realize a room-temperature ladder-type atomic vapor memory that operates on the Cs D1 line at 895\,nm. We provide a detailed experimental characterization and demonstration of on-demand storage and retrieval of weak coherent laser pulses with an average of 0.06(2) photons per pulse at a high signal-to-noise ratio of SNR$=830(80)$. The memory achieves a maximum internal efficiency of $\eta_{\text{int}}=15(1)\%$ and an estimated $1/e$-storage time of $\tau_{\mathrm{s}}\approx32\,$ns.
Benchmark properties for the storage of single photons from inhomogeneously broadened state-of-the-art solid-state emitters are estimated from the performance of the memory. Together with the immediate availability of high-quality InGaAs quantum dots emitting at 895\,nm, these results provide clear prospects for the development of 
a heterogeneous on-demand quantum light interface.

\end{abstract}

\maketitle

\section{\label{sec:intro}Introduction}

Quantum communication with single photons as flying qubits requires both quantum light sources and control over the timing and synchronization of photons. 
The ability of optical quantum memories to store and retrieve photons on demand~\cite{Lvovsky:2009, Hammerer:2010, Novikova:2012, Heshami:2016} allows for synchronization of probabilistic photons~\cite{Nunn:2013, Davidson:2023}, buffering and entanglement swapping in quantum repeater networks~\cite{Briegel:1998, Luo:2022, vanLeent:2022, Lei:2023}. A coherent light-matter interface is thus an integral part of future photonic quantum communication networks~\cite{Hammerer:2010,Chen:2021}. %
Technological applications will benefit from high bandwidths on the order of GHz and large photon 
rates, while maintaining low noise and high end-to-end memory efficiencies~\cite{Lvovsky:2009, Reim:2010, Hammerer:2010, Heshami:2016, Chen:2021}. 
A prominent host matter for an optical quantum memory is room-temperature alkali metal vapor contained in an evacuated glass cell~\cite{Lvovsky:2009, Hammerer:2010, Heshami:2016,KaiShinbrough.2023}.  
Large-scale and ~\cite{MeSSner:23} field applications~\cite{Wang2022}, and air- and spaceborne operations of such memory devices~\cite{Wallnoefer:2022, Guendogan:2021} are under investigation, enabled by the platform's technological simplicity 
compared to laser-cooled or cryogenic platforms. 

In room-temperature alkali vapors, coherent optical control over the light-matter interaction has been implemented in various atomic three-level systems, %
notably in $\Lambda$-type and ladder-type~($\Xi$) configurations. $\Lambda$-type memories can provide hundreds of milliseconds storage time~\cite{Katz.2018}, but are typically prone to readout noise and true single-photon operation has been reached in only a few experiments~\cite{Buser:2023,Dideriksen.2021}. Because of the large energy separation between the initial and the storage states the $\Xi$-type configurations are superior in this regard and provide storage times up to about 100\,ns~\cite{Davidson:2023-2}.

In ladder-type level configurations, GHz-bandwidth heralded single photons at 852\,nm have been stored by two-photon off-resonant cascaded absorption (ORCA) on the D2 line of Cs~\cite{Kaczmarek:2018}. 
A fast ladder memory (FLAME) on the D2 line of Rb~\cite{Finkelstein:2018, Davidson:2023-2} 
enabled synchronization of single photons at 780\,nm  stemming from identical, and thus intrinsically well-matched, atomic vapors, resulting in an enhanced photon coincidence rate~\cite{Davidson:2023}. 
Moreover, the ORCA protocol in Rb has enabled a telecom-C-band quantum memory~\cite{Thomas:2023}. %

Semiconductor quantum dot (QD) single-photon sources can emit deterministically, with high brightness and low multiphoton contribution~\cite{Gschrey.2015}, and are excellent building blocks for photonic quantum information technology~\cite{Heindel:2023,Uppu:2021}. 
Self-assembled InGaAs QDs typically exhibit an exciton lifetime of about 1\,ns and the emission wavelength can be tuned from NIR~\cite{Wang.2019} up to the telecom-C-band~\cite{Nawrath.2023} by engineering the epitaxial growth conditions and the material composition. This renders InGaAs QDs particularly interesting in the context of this work because they can be aligned to the memory's operational wavelength of 895\,nm. Embedding such QDs into an optical cavity, single photons have been generated on demand with a probability of up to 57\% and with an average two-photon interference visibility of 97.5\% at the output of the final optical fiber~\cite{Tomm.2021}.
Due to this outstanding performance it has been a long-lasting goal to couple InGaAs QDs to atomic quantum memories.
Along this route, atomic vapor was used as spectrally selective and tunable delay line for quantum dot 
photons~\cite{Wildmann:2015,Bremer:2020, Vural:2021}.
Also, quantum dot photons were delayed by 5\,ns using the strong dispersion between two hyperfine-split excited state transitions of the Cs D1 line~\cite{Kroh:2019}. %
More recently, a coherent exchange of quanta between single photons from a GaAs quantum dot and $^{87}$Rb vapor has been shown~\cite{Cui:2023}, and photons from quantum dots embedded in semiconductor nanowires have been slowed down in an atomic vapor~\cite{AlMaruf:2023}.
Moreover, a bandwidth-matched $\Lambda$-type memory has been demonstrated~\cite{Wolters:2017}. 
Recently, a landmark experiment demonstrated deterministic storage and retrieval of photons from an InAs QD emitting at 1529\,nm~\cite{Thomas:2024}. However, the characteristic storage time of the memory was fundamentally limited to 1.1\,ns, only a factor of 1.3 above the QD exciton lifetime.

Here, we present a ladder-type memory on the Cs D1 line, allowing for low-noise storage and on-demand retrieval of high-bandwidth photons at 895\,nm, directly compatible with quantum-dot-based single-photon sources from Refs.~\cite{Wildmann:2015,Kroh:2019,Wang.2019, Bremer:2020, Vural:2021}. The memory exhibits a $1/e$-storage time $\tau_\text{s}$ of tens of nanoseconds, an order of magnitude longer than typical exciton lifetimes. 
These render the memory ideal for the storage of single photons from QDs with real-time electro-optic memory control.

\section{Memory protocol}
\label{sec:protocol}

We operate the memory device with signal photons slightly red-detuned from the D1 line of Cs, as depicted in Fig.~\ref{fig:setup}a. 
A strong control field at a wavelength of 876\,nm couples the intermediate 6P$_{1/2}$ level to the 6D$_{3/2}$ level. The memory operates in the crossover regime between the electromagnetically induced transparency (EIT) and Autler-Townes (ATS) protocols~\cite{Rastogi:2019}. 

Pulsed operation maps the signal field onto a collective orbital coherence of the cesium atoms between the ground state and the storage level
6D$_{3/2}$, the spinwave. 

Re-applying the control pulse coherently retrieves the initial signal photon.

In this memory scheme, the main source of decoherence of the 
collective excitation is motional dephasing~\cite{Main:2021}. The wavevector mismatch of the signal and control fields determines the wavelength of the spinwave, here $43.5\,\mu$m, and therewith the time scale of the spinwave decay. This relation is depicted in Fig.~\ref{fig:storagetimescan} and further discussed in Sec.~\ref{subsec:optimisation}. In ladder-type memories the storage state is effectively unpopulated in thermal equilibrium, and the frequency mismatch between the signal and control transitions is very large compared to the Doppler-broadened linewidths. These two facts lead to extremely high signal-to-noise ratios (SNR). The storage time of the two-photon ladder-type memory is ultimately limited by the lifetime of the employed upper state. It has been shown that, in principle, the limitations of motional dephasing can be mitigated by a continuous protection mechanism using additional off-resonant optical dressing~\cite{Davidson:2023}, or by velocity-selective pumping~\cite{Main:2021} at the cost of reduced optical depth.

\section{\label{sec:exp}Experimental setup}

\begin{figure*}
\centering
\includegraphics{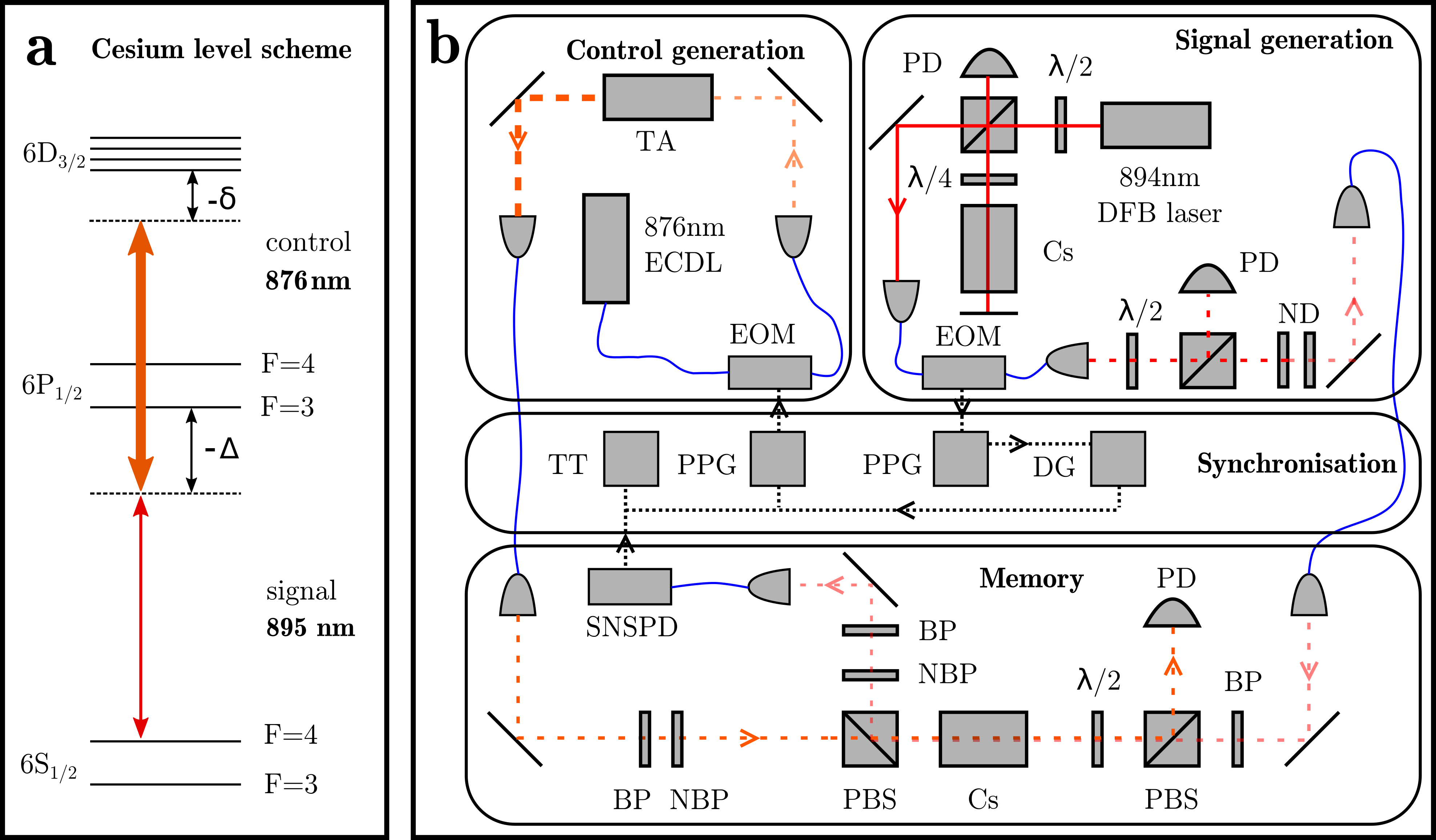}
  \caption{\label{fig:setup}
  Overview of the experiment. (a)~Reduced level scheme of Cs. Signal and control lasers in a ladder-type configuration with single-photon detuning $\Delta$ and two-photon detuning $\delta$. (b)~Experimental setup. Control pulses are generated with an electro-optical amplitude modulator (EOM) and subsequently amplified by a tapered amplifier (TA). Spectral filtering is achieved with a bandpass filter (BP) and a narrow-bandpass filter (NBP). A Doppler-free cesium (Cs) spectroscopy serves as a frequency reference for the signal laser. Signal pulses are generated with an EOM and monitored using a fast photodiode (PD). The signal beam is attenuated to single-photon level with neutral-density filters (ND). Counterpropagating signal and control pulses overlap inside the heated vapor cell (Cs). Signal photons are detected with a superconducting nanowire single-photon detector (SNSPD). Control leakage is suppressed by spatial, polarization and spectral filtering (BP, NBP). Programmable pulse generators (PPG) drive the EOMs and are synchronized with a digital delay generator (DG).} 

\end{figure*}

The setup as depicted in Fig.~\ref{fig:setup}b consists of the signal generation, the control generation and the memory unit. We describe the individual parts and their synchronization in the next sections.

\subsection{Cs Cell and Heating System}

The cylindrical vapor cell is 25\,mm in diameter and 25\,mm in length. The requirement for a homogeneous optical field in the interaction area limits the Rayleigh range to the length of the cell and thereby the beam waists of signal and control lasers as well. In this experiment we chose a signal beam diameter of $100(5)\,\mu$m and a control beam diameter of $145(5)\,\mu$m at the center of the glass cell. The beam waists were measured using a CMOS camera with a pixel size of 5\,µm. Since the memory protocol is faster than the time scale of decoherence due to cesium-cesium collisions ($\approx$\,$150\mu$s~\cite{DJCroucher.1969}) the experiment does not require a buffer-gas filled cell.
A customized heating system made of a spring-loaded copper block with a heater allows for uniform heating of the glass cell to a temperature of up to 120°C while preventing condensation of Cs on the entrance windows.

\subsection{Lasers, Pulse Generation and Timing Controls}

The signal beam is generated by a continuous-wave (CW) 895\,nm distributed-feedback laser diode (DFB). A fraction of the beam is used for Doppler-free spectroscopy, which serves, in addition to a wavemeter with 100\,MHz resolution, as a wavelength reference. An electro-optical amplitude modulator (EOM) generates signal pulses with a rise time of 500\,ps. After attenuation to single-photon level with neutral-density (ND) filters, the signal beam is directed into the memory cell. 
The control beam is generated by a CW 876\,nm external-cavity diode laser (ECDL) with an output power of up to 50\,mW. The control pulses are carved 
 from the CW laser with an EOM and subsequently amplified in a tapered amplifier (TA). The high-power output of the TA is then frequency-filtered with a 10\,nm bandpass filter and a 1\,nm narrowband filter to remove the amplified spontaneous emission (ASE) background emerging from the TA. Pulse shaping with the EOMs allows for 1\,ns-long (FWHM) pulses. The electronic control signals for the EOMs are generated with a 512\,bit programmable pulse generator (PPG) with a repetition rate of 10\,MHz and a resolution of $200$\,ps. The pulses are synchronized using a digital delay generator with 10\,ps resolution. The shape of the pulses and the extinction of the EOMs are monitored on additional photodiodes. The setup allows for variable adjustment of pulse lengths and timings of both the signal and control lasers independently. The repetition rate of the experiment can easily be matched to external single-photon sources. 

\subsection{Signal Filtering and Detection}
The counterpropagating and orthogonally polarized signal and control pulses overlap inside the cesium cell. The control pulses are separated in polarization and monitored on a fast 
photodiode. After being stored in the memory cell, the transmitted signal is also polarization-selected. After a series of broadband filters and a 1\,nm narrow-band interference filter, the transmitted signal is coupled into a polarization-maintaining optical fiber that is connected to a superconducting nanowire single-photon detector (SNSPD) with a nominal detection efficiency of 0.85 and a temporal resolution of 50\,ps.

\section{Characterization and Results}
\label{sec:results}

In this section we characterize the memory and evaluate its performance regarding storage time and efficiency. Additionally, we determine the spectral acceptance window of the memory and use the results to benchmark the memory performance for the feasibility of storing single photons from realistic quantum dot devices.

\subsection{Optimization of Efficiency and Storage Time}
\label{subsec:optimisation}

\begin{figure}[h]
\includegraphics[width=\columnwidth]{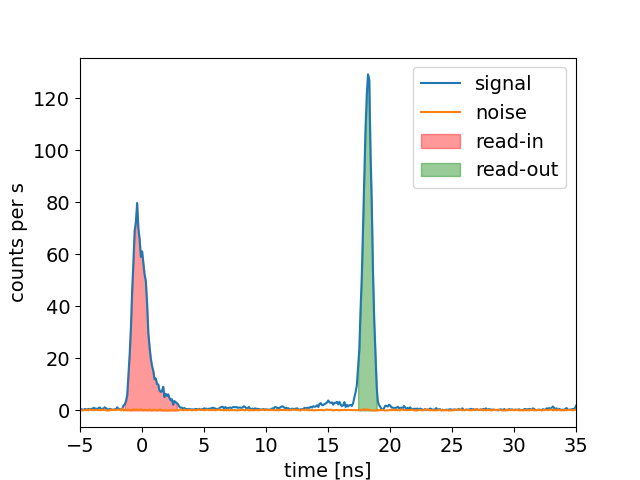}%
 \caption{
  Exemplary storage trace. Histogrammed 
  storage and retrieval of 800\,ps FWHM-long signal pulses attenuated to 0.06(2) photons per pulse at -200\,MHz single-photon detuning from the $F=4$ to $F'=3$ transition of the D1 line. The control pulses have a FWHM of 1\,ns and yield a peak Rabi frequency of $\Omega=2\pi\cdot 900(30)$\,MHz. At the retrieval time of 17.4(1)\,ns, the memory achieves an internal efficiency of $\eta_{\text{int}}=15(1)\%$. The data were collected for an integration time of 10\,s with a bin size of 100\,ps. The shaded areas mark the integration windows for the efficiency calculations.}
\label{fig:storagetrace}
\end{figure}

Fig.\,\ref{fig:storagetrace} shows the histogrammed result of a typical storage and retrieval experiment. The signal was attenuated to an average of 0.06(2) photons per pulse, and the integration time was 10\,s using time bins of 100\,ps. We use these parameters in all measurements unless stated otherwise.  
Leakage due to imperfect storage of incoming photons produces the read-in peak (shaded red). This effect can be attributed to imperfect mapping between the signal field and the spinwave. The photons in the read-out peak (shaded green) are retrieved from the atomic ensemble by the second control laser pulse after a delay of $17.4(1)$\,ns. The shaded areas mark the integration windows for the efficiency analysis which capture over 90\% of the retrieved signal while keeping the SNR high~\cite{Esguerra:2023}. The noise curve is taken with the input signal blocked, i.e.\ it
contains both the leaked control laser light and the integrated dark counts from the detector. The noise floor stems from insufficient suppression by the EOMs, electrical noise from the pulse generation, and EOM bias drifts. A far-detuned ($\sim6$\,GHz) signal measurement without the control laser serves as a reference for the signal input. We calculate the internal memory efficiency as the ratio of integrated counts within the retrieved signal peak and the reference measurement, $\eta_{\text{int}} = N_{\text{ret}} / N_{\text{in}}$, and find a maximum value of $\eta_{\text{int}}=15(1)\%$ at 17.4(2)\,ns storage time. Including the setup transmission ($T$), we achieve an end-to-end memory efficiency of $\eta_\mathrm{e2e}=\eta_{\text{int}}\cdot T=2.2(2)\%$. We expect an increased end-to-end efficiency by straightforward improvements of the setup transmission, e.g. using an anti-reflective coated Cs cell in future experiments. In a separate experiment with long integration time, the low noise floor and an SNR of $830(80)$ are confirmed. 

\begin{figure}[h]
\includegraphics[width=\columnwidth]{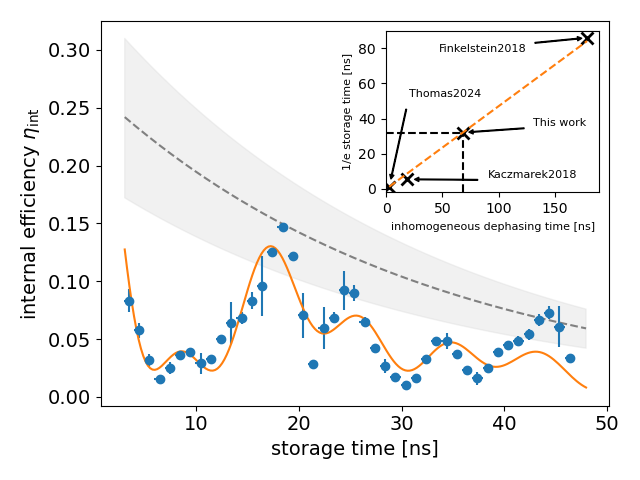}
\caption{Measured dynamics of the internal efficiency. In every step, the pulse timings are optimized. The peak Rabi frequency is estimated to $2\pi \cdot 900(30)$\,MHz.
Errors represent fit uncertainties and time resolution of the pulse generator. Integration times long compared to the inverse repetition rate render statistical errors negligible w.r.t.\ fit errors. The inset shows the relation between the inhomogeneous dephasing time and the $1/e$-storage time as reported in the literature (Kaczmarek2018~\cite{Kaczmarek:2018}, Finkelstein2018~\cite{Finkelstein:2018}, Thomas2024~\cite{Thomas:2024}). The gray-shaded area estimates the exponential envelope of the spinwave oscillations.
}
\label{fig:storagetimescan}
\end{figure}

\begin{figure}[h]
\includegraphics[width=\columnwidth]{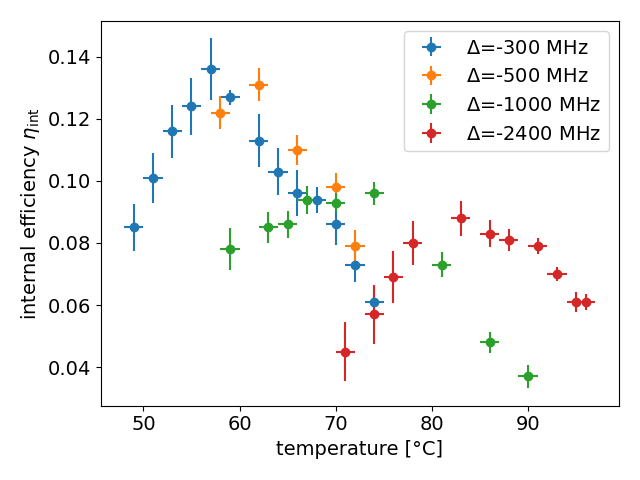}
\caption{\label{fig:OD}
Temperature dependence of the optimal working point of the memory. The OD is increased by raising the temperature of the cell. The larger the signal detuning $\Delta$, the more OD is required to achieve the highest internal efficiency. For large single-photon detunings $\Delta < -1000$\,MHz, the efficiency is limited by the available control power. 
}
\end{figure}

The measured dynamics of the internal efficiency are displayed in Fig.~\ref{fig:storagetimescan}. We scan the storage time $\tau_\mathrm{s}$ by adapting the timing of the retrieval control pulse. The memory is operated at 200\,MHz single-photon red-detuning and the cell temperature is optimized, as discussed later and presented in Fig.\,\ref{fig:OD}.
The peak Rabi frequency of the Gaussian control pulses is estimated to be $2\pi \cdot 900(30)$\,MHz, given the measured CW laser power and the beam geometry as described in Sec.\,\ref{sec:exp}.
Retrieval times $\tau_\mathrm{s}<3$\,ns are inaccessible because of otherwise temporally overlapping read-in and read-out control pulses. 
The data shows strong oscillatory spinwave dynamics that can be attributed to multiple addressed hyperfine states of the 6D$_{3/2}$ storage level. This results in a beating of the generated spinwaves with the frequency splittings within the hyperfine manifold. It has been shown that this beating can be suppressed if only one hyperfine state is addressed by the signal field when circularly pumping into a stretched state~\cite{Finkelstein:2018}. 
The model introduced in Ref.\,\cite{Finkelstein:2018} (solid line) qualitatively reproduces the data. The inhomogeneous dephasing time of the spinwave $T_\text{inhom}$ depends on its wavelength $\lambda_\text{s}$, determined by the wavevector mismatch of signal and control laser, and the thermal velocity of the atoms $v_\text{th}$: $T_\text{inhom}=\lambda_\text{s}/(\sqrt{2}\pi v_\text{th})$~\cite{Finkelstein:2021}. The inset of Fig.\,\ref{fig:storagetimescan} depicts the relation between the theoretical dephasing times and the measured $1/e$-storage times that have been reported in previously ladder-type memories~\cite{Finkelstein:2018, Kaczmarek:2018, Thomas:2023}. %
For the level configuration in Fig.~\ref{fig:setup}a, we expect a $1/e$-storage time of about $\tau_\mathrm{s}\approx32$\,ns, which is consistent with the data. The gray-dashed line and the shaded area estimate the exponential envelope of the spinwave oscillations. The zero-time efficiency cannot be faithfully determined from the data. 
All following data are acquired at the maximum efficiency retrieval time of 17.4\,ns. Future experiments will also shed light on additional inhomogeneous decay mechanisms such as spin-exchange collisions or atoms leaving the interaction area.

The influence of the cell temperature on the internal memory efficiency is presented in Fig.\,\ref{fig:OD}. We find an optimum working point that shifts to higher temperatures for larger signal detunings $\Delta$. The theory for optimal light storage~\cite{GorshkovPhysRevA.76.033806} suggests that the internal memory efficiency increases monotonically with the optical depth (OD) of the atomic vapor. In agreement with other experiments~\cite{Dujic.2024, Phillips:2008}, we observe a decrease in efficiency  for a given single-photon detuning if the OD is too high. Additional leakage channels, such as four-wave mixing and Raman scattering processes are suggested as possible sources for the observed decrease. However, this cannot hold as the explanation in this case because the two-color ladder-type configuration inhibits these types of mechanisms. More likely, the efficiencies in the experiment are limited by the Rabi frequency of $2\pi\cdot900(30)$\,MHz. The role of the control pulse shape should be the subject of further investigations. Here, we have constrained the data evaluation to red-detuned signal input because the efficiencies are much lower in the blue-detuned regime. We attribute this behavior to additional spinwave dephasing due to increased coupling to the $F^\prime = 4$ intermediate state.

\subsection{Spectral Acceptance Window}
\label{subsec:bandwidth}

Solid-state single-photon sources such as self-assembled quantum dots usually suffer from an inhomogeneous broadening of the emission linewidth due to spectral diffusion~\cite{Thoma:2016}. It is thus important to quantify the spectral region in which photons of a given bandwidth can be stored efficiently.
To estimate this spectral acceptance window of the memory $\mathcal{W}$, we measure the internal efficiency over a scan of the two-photon detuning $\delta$ for a set of signal detunings, $\Delta = \{-300,-500,-1000\}$\,MHz.
This data set emulates photon emission from spectrally broadened quantum light sources. %

\begin{figure}[h]
\includegraphics[width=\columnwidth]{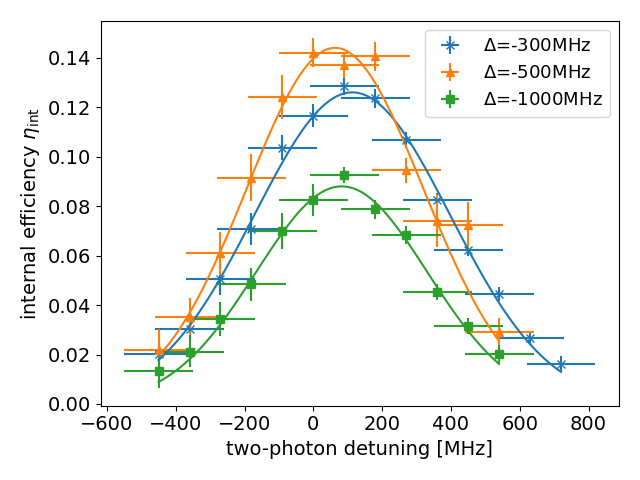} %
\caption{\label{fig:bandwidth} %
Spectral acceptance window. Internal memory efficiency over two-photon detuning for different single-photon detunings and 1\,ns signal pulse lengths. We scan the signal frequency keeping the control frequency fixed. The width of the curve corresponds to the spectral acceptance window of the memory convoluted with the input laser pulse. For $\Delta=-500$~MHz we extract $\mathcal{W}=560(60)~$MHz.
}
\end{figure}

Taking the Fourier-limited linewidth of the signal laser, $\gamma_\mathrm{hom}\approx 440$\,MHz, into account, we can infer the spectral acceptance window of the memory from the data presented in Fig.~\ref{fig:bandwidth}. 

We use this result to estimate the influence of inhomogeneously broadened signals on the internal memory efficiency and SNR in the next section. 

\subsection{Benchmarks for Storage of Single Photons}
\label{subsec:benchmark}

By virtue of the spectral acceptance window we can estimate benchmark properties for true single-photon input that are required for successful storage experiments.
We assume that the signal input is red-detuned from the signal transition, $\Delta=-500$\,MHz, and that the memory acceptance window has a FWHM of $\mathcal{W}=560(60)$\,MHz. From that, we can infer the percentage of emitted quantum dot photons that can be stored and retrieved. Considering the expected source efficiency, i.e.\ the probability of extracting a single photon upon excitation $\eta_{src}$, the fiber-coupled QD setup transmission $\eta_{\text{T}}=0.01$ and the polarization losses due filtering of a single exciton transition $\eta_{P}=0.5$, we estimate the effective count rate of a quantum dot light source.  Assuming a constant noise floor that is dominated by control field leakage we can find values for the expected signal-to-noise ratio in a quantum dot--memory interface. We account for the temporal mode mismatch between the quantum dot photon and the signal laser by considering the benchmarking results from \cite{Gao:2019}, in particular the effective overlap between the dominant quantum dot mode and the memory mode $\eta_{\text{QD-mem}}=0.66$.\\

\begin{figure}[h]
\includegraphics[width=\columnwidth]{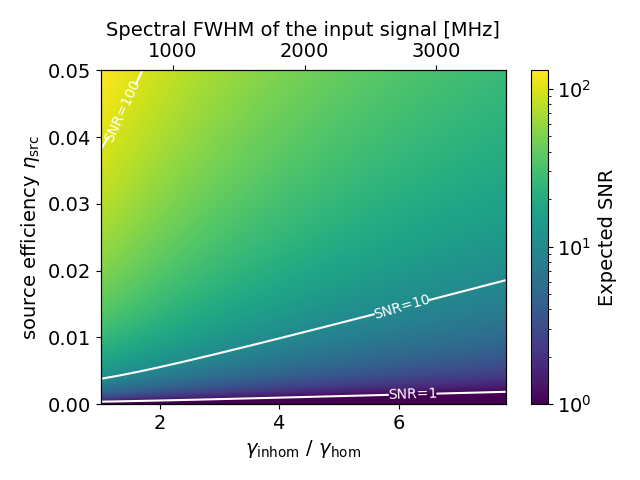}
\caption{\label{fig:bechmark} %
Benchmarking the memory for single photons from a semiconductor quantum dot source with 400\,MHz homogeneous linewidth. %
Influence of the photon extraction efficiency of the light source and the inhomogeneous linewidth $\gamma_\mathrm{inhom}$ of the emitted photons on the expected SNR of the storage signal.}
\end{figure}

Figure\,\ref{fig:bechmark} shows the estimated upper bound on the expected SNRs for different source efficiencies and inhomogeneous linewidths $\gamma_\mathrm{inhom}$ of the input photons. Considering a typical InGaAs quantum dot sample \cite{Bremer:2020}, we expect single-photon emission with a pulse length of about 1\,ns and a homogeneous linewidth of about 400\,MHz which fits into the acceptance window of the memory.
Even in the case of inhomogeneously broadened emission to a linewidth of about 2\,GHz due to spectral diffusion we expect SNRs above 10. We want to point out that the noise background in an experiment will be lowered further by unfiltered background emission, i.e. from undesired excitation of multiple emitters or collection of light from additional relaxation paths, e.g. bi-excitons.

\section{Discussion}
\label{sec:discussion}
We have demonstrated on-demand storage and retrieval of attenuated laser pulses with an average photon number of 0.06(2) in a room-temperature vapor ladder-type memory on the Cs D1 line. The memory achieves a maximum internal efficiency of $\eta_\mathrm{int}=15(1)\%$ and an end-to-end efficiency of $\eta_\mathrm{e2e}=2.2(2)\%$. The high SNR of 830(80) for ultraweak signals and the broad spectral acceptance window of $\mathcal{W}=560(60)$\,MHz promise compatibility of the memory with state-of-the-art solid-state single-photon emitters such as InGaAs quantum dots~\cite{Wang.2019}. The variable control and signal pulse lengths, timings and pulse shapes in the experiment offer unprecedented flexibility when coupling the quantum dot single-photon source to the memory setup.

\section{Outlook}
\label{sec:outlook}
We see clear technological improvements that will increase the efficiency of the optical memory in future experiments. The implementation of optical hyperfine pumping will significantly boost the memory performance. The use of an antireflection-coated cell will further increase the overall setup transmission and thereby the end-to-end efficiency. An active stabilization of the EOMs' DC-bias drifts will improve the pulse quality and the long-term stability of the experiment. To obtain a more homogeneous control field along the signal beam profile, we will aim at a larger difference in beam diameters in future experiments.

The results show that the memory is compatible with state-of-the-art quantum light emitters. The interface of an external single-photon source with the memory poses a significant technological challenge, especially regarding wavelength matching, bandwidth, spectral drifts of the source, and synchronization. 
The aforementioned setup improvements will further facilitate the construction of a heterogeneous on-demand light--matter interface. The study of such systems may contribute to the development of optical quantum networks. 

\begin{acknowledgments}
We would like to thank Omri Davidson, Eilon Poem, and Ofer Firstenberg for fruitful discussions and Ilja Gerhardt for providing the vapor cell.
This work was funded by the DFG project \mbox{RE2974/28-1} (\mbox{448532670}) and the Federal Ministry of Education and Research (BMBF) project Q-ToRX \mbox{16KISQ040K}.
\end{acknowledgments}

\bibliography{references}

\begin{thebibliography}{48}%
\makeatletter
\providecommand \@ifxundefined [1]{%
 \@ifx{#1\undefined}
}%
\providecommand \@ifnum [1]{%
 \ifnum #1\expandafter \@firstoftwo
 \else \expandafter \@secondoftwo
 \fi
}%
\providecommand \@ifx [1]{%
 \ifx #1\expandafter \@firstoftwo
 \else \expandafter \@secondoftwo
 \fi
}%
\providecommand \natexlab [1]{#1}%
\providecommand \enquote  [1]{``#1''}%
\providecommand \bibnamefont  [1]{#1}%
\providecommand \bibfnamefont [1]{#1}%
\providecommand \citenamefont [1]{#1}%
\providecommand \href@noop [0]{\@secondoftwo}%
\providecommand \href [0]{\begingroup \@sanitize@url \@href}%
\providecommand \@href[1]{\@@startlink{#1}\@@href}%
\providecommand \@@href[1]{\endgroup#1\@@endlink}%
\providecommand \@sanitize@url [0]{\catcode `\\12\catcode `\$12\catcode `\&12\catcode `\#12\catcode `\^12\catcode `\_12\catcode `\%12\relax}%
\providecommand \@@startlink[1]{}%
\providecommand \@@endlink[0]{}%
\providecommand \url  [0]{\begingroup\@sanitize@url \@url }%
\providecommand \@url [1]{\endgroup\@href {#1}{\urlprefix }}%
\providecommand \urlprefix  [0]{URL }%
\providecommand \Eprint [0]{\href }%
\providecommand \doibase [0]{https://doi.org/}%
\providecommand \selectlanguage [0]{\@gobble}%
\providecommand \bibinfo  [0]{\@secondoftwo}%
\providecommand \bibfield  [0]{\@secondoftwo}%
\providecommand \translation [1]{[#1]}%
\providecommand \BibitemOpen [0]{}%
\providecommand \bibitemStop [0]{}%
\providecommand \bibitemNoStop [0]{.\EOS\space}%
\providecommand \EOS [0]{\spacefactor3000\relax}%
\providecommand \BibitemShut  [1]{\csname bibitem#1\endcsname}%
\let\auto@bib@innerbib\@empty
\bibitem [{\citenamefont {Lvovsky}\ \emph {et~al.}(2009)\citenamefont {Lvovsky}, \citenamefont {Sanders},\ and\ \citenamefont {Tittel}}]{Lvovsky:2009}%
  \BibitemOpen
  \bibfield  {author} {\bibinfo {author} {\bibfnamefont {A.~I.}\ \bibnamefont {Lvovsky}}, \bibinfo {author} {\bibfnamefont {B.~C.}\ \bibnamefont {Sanders}},\ and\ \bibinfo {author} {\bibfnamefont {W.}~\bibnamefont {Tittel}},\ }\bibfield  {title} {\bibinfo {title} {{Optical quantum memory}},\ }\href {https://doi.org/https://doi.org/10.1038/nphoton.2009.231} {\bibfield  {journal} {\bibinfo  {journal} {Nature Photonics}\ }\textbf {\bibinfo {volume} {3}},\ \bibinfo {pages} {706} (\bibinfo {year} {2009})}\BibitemShut {NoStop}%
\bibitem [{\citenamefont {Hammerer}\ \emph {et~al.}(2010)\citenamefont {Hammerer}, \citenamefont {S\o{}rensen},\ and\ \citenamefont {Polzik}}]{Hammerer:2010}%
  \BibitemOpen
  \bibfield  {author} {\bibinfo {author} {\bibfnamefont {K.}~\bibnamefont {Hammerer}}, \bibinfo {author} {\bibfnamefont {A.~S.}\ \bibnamefont {S\o{}rensen}},\ and\ \bibinfo {author} {\bibfnamefont {E.~S.}\ \bibnamefont {Polzik}},\ }\bibfield  {title} {\bibinfo {title} {Quantum interface between light and atomic ensembles},\ }\href {https://doi.org/10.1103/RevModPhys.82.1041} {\bibfield  {journal} {\bibinfo  {journal} {Rev. Mod. Phys.}\ }\textbf {\bibinfo {volume} {82}},\ \bibinfo {pages} {1041} (\bibinfo {year} {2010})}\BibitemShut {NoStop}%
\bibitem [{\citenamefont {Novikova}\ \emph {et~al.}(2012)\citenamefont {Novikova}, \citenamefont {Walsworth},\ and\ \citenamefont {Xiao}}]{Novikova:2012}%
  \BibitemOpen
  \bibfield  {author} {\bibinfo {author} {\bibfnamefont {I.}~\bibnamefont {Novikova}}, \bibinfo {author} {\bibfnamefont {R.}~\bibnamefont {Walsworth}},\ and\ \bibinfo {author} {\bibfnamefont {Y.}~\bibnamefont {Xiao}},\ }\bibfield  {title} {\bibinfo {title} {Electromagnetically induced transparency-based slow and stored light in warm atoms},\ }\href {https://doi.org/https://doi.org/10.1002/lpor.201100021} {\bibfield  {journal} {\bibinfo  {journal} {Laser \& Photonics Reviews}\ }\textbf {\bibinfo {volume} {6}},\ \bibinfo {pages} {333} (\bibinfo {year} {2012})}\BibitemShut {NoStop}%
\bibitem [{\citenamefont {Heshami}\ \emph {et~al.}(2016)\citenamefont {Heshami}, \citenamefont {England}, \citenamefont {Humphreys}, \citenamefont {Bustard}, \citenamefont {Acosta}, \citenamefont {Nunn},\ and\ \citenamefont {Sussman}}]{Heshami:2016}%
  \BibitemOpen
  \bibfield  {author} {\bibinfo {author} {\bibfnamefont {K.}~\bibnamefont {Heshami}}, \bibinfo {author} {\bibfnamefont {D.~G.}\ \bibnamefont {England}}, \bibinfo {author} {\bibfnamefont {P.~C.}\ \bibnamefont {Humphreys}}, \bibinfo {author} {\bibfnamefont {P.~J.}\ \bibnamefont {Bustard}}, \bibinfo {author} {\bibfnamefont {V.~M.}\ \bibnamefont {Acosta}}, \bibinfo {author} {\bibfnamefont {J.}~\bibnamefont {Nunn}},\ and\ \bibinfo {author} {\bibfnamefont {B.~J.}\ \bibnamefont {Sussman}},\ }\bibfield  {title} {\bibinfo {title} {Quantum memories: emerging applications and recent advances},\ }\href {https://doi.org/10.1080/09500340.2016.1148212} {\bibfield  {journal} {\bibinfo  {journal} {Journal of Modern Optics}\ }\textbf {\bibinfo {volume} {63}},\ \bibinfo {pages} {2005} (\bibinfo {year} {2016})}\BibitemShut {NoStop}%
\bibitem [{\citenamefont {Nunn}\ \emph {et~al.}(2013)\citenamefont {Nunn}, \citenamefont {Langford}, \citenamefont {Kolthammer}, \citenamefont {Champion}, \citenamefont {Sprague}, \citenamefont {Michelberger}, \citenamefont {Jin}, \citenamefont {England},\ and\ \citenamefont {Walmsley}}]{Nunn:2013}%
  \BibitemOpen
  \bibfield  {author} {\bibinfo {author} {\bibfnamefont {J.}~\bibnamefont {Nunn}}, \bibinfo {author} {\bibfnamefont {N.~K.}\ \bibnamefont {Langford}}, \bibinfo {author} {\bibfnamefont {W.~S.}\ \bibnamefont {Kolthammer}}, \bibinfo {author} {\bibfnamefont {T.~F.~M.}\ \bibnamefont {Champion}}, \bibinfo {author} {\bibfnamefont {M.~R.}\ \bibnamefont {Sprague}}, \bibinfo {author} {\bibfnamefont {P.~S.}\ \bibnamefont {Michelberger}}, \bibinfo {author} {\bibfnamefont {X.-M.}\ \bibnamefont {Jin}}, \bibinfo {author} {\bibfnamefont {D.~G.}\ \bibnamefont {England}},\ and\ \bibinfo {author} {\bibfnamefont {I.~A.}\ \bibnamefont {Walmsley}},\ }\bibfield  {title} {\bibinfo {title} {Enhancing multiphoton rates with quantum memories},\ }\href {https://doi.org/10.1103/PhysRevLett.110.133601} {\bibfield  {journal} {\bibinfo  {journal} {Phys. Rev. Lett.}\ }\textbf {\bibinfo {volume} {110}},\ \bibinfo {pages} {133601} (\bibinfo {year} {2013})}\BibitemShut {NoStop}%
\bibitem [{\citenamefont {Davidson}\ \emph {et~al.}(2023{\natexlab{a}})\citenamefont {Davidson}, \citenamefont {Yogev}, \citenamefont {Poem},\ and\ \citenamefont {Firstenberg}}]{Davidson:2023}%
  \BibitemOpen
  \bibfield  {author} {\bibinfo {author} {\bibfnamefont {O.}~\bibnamefont {Davidson}}, \bibinfo {author} {\bibfnamefont {O.}~\bibnamefont {Yogev}}, \bibinfo {author} {\bibfnamefont {E.}~\bibnamefont {Poem}},\ and\ \bibinfo {author} {\bibfnamefont {O.}~\bibnamefont {Firstenberg}},\ }\bibfield  {title} {\bibinfo {title} {Single-photon synchronization with a room-temperature atomic quantum memory},\ }\href {https://doi.org/10.1103/PhysRevLett.131.033601} {\bibfield  {journal} {\bibinfo  {journal} {Phys. Rev. Lett.}\ }\textbf {\bibinfo {volume} {131}},\ \bibinfo {pages} {033601} (\bibinfo {year} {2023}{\natexlab{a}})}\BibitemShut {NoStop}%
\bibitem [{\citenamefont {Briegel}\ \emph {et~al.}(1998)\citenamefont {Briegel}, \citenamefont {D\"ur}, \citenamefont {Cirac},\ and\ \citenamefont {Zoller}}]{Briegel:1998}%
  \BibitemOpen
  \bibfield  {author} {\bibinfo {author} {\bibfnamefont {H.-J.}\ \bibnamefont {Briegel}}, \bibinfo {author} {\bibfnamefont {W.}~\bibnamefont {D\"ur}}, \bibinfo {author} {\bibfnamefont {J.~I.}\ \bibnamefont {Cirac}},\ and\ \bibinfo {author} {\bibfnamefont {P.}~\bibnamefont {Zoller}},\ }\bibfield  {title} {\bibinfo {title} {Quantum repeaters: The role of imperfect local operations in quantum communication},\ }\href {https://doi.org/10.1103/PhysRevLett.81.5932} {\bibfield  {journal} {\bibinfo  {journal} {Phys. Rev. Lett.}\ }\textbf {\bibinfo {volume} {81}},\ \bibinfo {pages} {5932} (\bibinfo {year} {1998})}\BibitemShut {NoStop}%
\bibitem [{\citenamefont {Luo}\ \emph {et~al.}(2022)\citenamefont {Luo}, \citenamefont {Yu}, \citenamefont {Liu}, \citenamefont {Zheng}, \citenamefont {Wang}, \citenamefont {Wang}, \citenamefont {Li}, \citenamefont {Jiang}, \citenamefont {Xie}, \citenamefont {Zhang}, \citenamefont {Bao},\ and\ \citenamefont {Pan}}]{Luo:2022}%
  \BibitemOpen
  \bibfield  {author} {\bibinfo {author} {\bibfnamefont {X.-Y.}\ \bibnamefont {Luo}}, \bibinfo {author} {\bibfnamefont {Y.}~\bibnamefont {Yu}}, \bibinfo {author} {\bibfnamefont {J.-L.}\ \bibnamefont {Liu}}, \bibinfo {author} {\bibfnamefont {M.-Y.}\ \bibnamefont {Zheng}}, \bibinfo {author} {\bibfnamefont {C.-Y.}\ \bibnamefont {Wang}}, \bibinfo {author} {\bibfnamefont {B.}~\bibnamefont {Wang}}, \bibinfo {author} {\bibfnamefont {J.}~\bibnamefont {Li}}, \bibinfo {author} {\bibfnamefont {X.}~\bibnamefont {Jiang}}, \bibinfo {author} {\bibfnamefont {X.-P.}\ \bibnamefont {Xie}}, \bibinfo {author} {\bibfnamefont {Q.}~\bibnamefont {Zhang}}, \bibinfo {author} {\bibfnamefont {X.-H.}\ \bibnamefont {Bao}},\ and\ \bibinfo {author} {\bibfnamefont {J.-W.}\ \bibnamefont {Pan}},\ }\bibfield  {title} {\bibinfo {title} {Postselected entanglement between two atomic ensembles separated by 12.5 km},\ }\href {https://doi.org/10.1103/PhysRevLett.129.050503} {\bibfield  {journal} {\bibinfo  {journal} {Phys. Rev. Lett.}\ }\textbf
  {\bibinfo {volume} {129}},\ \bibinfo {pages} {050503} (\bibinfo {year} {2022})}\BibitemShut {NoStop}%
\bibitem [{\citenamefont {van Leent}\ \emph {et~al.}(2022)\citenamefont {van Leent}, \citenamefont {Bock}, \citenamefont {Fertig}, \citenamefont {Garthoff}, \citenamefont {Eppelt}, \citenamefont {Zhou}, \citenamefont {Malik}, \citenamefont {Seubert}, \citenamefont {Bauer}, \citenamefont {Rosenfeld}, \citenamefont {Zhang}, \citenamefont {Becher},\ and\ \citenamefont {Weinfurter}}]{vanLeent:2022}%
  \BibitemOpen
  \bibfield  {author} {\bibinfo {author} {\bibfnamefont {T.}~\bibnamefont {van Leent}}, \bibinfo {author} {\bibfnamefont {M.}~\bibnamefont {Bock}}, \bibinfo {author} {\bibfnamefont {F.}~\bibnamefont {Fertig}}, \bibinfo {author} {\bibfnamefont {R.}~\bibnamefont {Garthoff}}, \bibinfo {author} {\bibfnamefont {S.}~\bibnamefont {Eppelt}}, \bibinfo {author} {\bibfnamefont {Y.}~\bibnamefont {Zhou}}, \bibinfo {author} {\bibfnamefont {P.}~\bibnamefont {Malik}}, \bibinfo {author} {\bibfnamefont {M.}~\bibnamefont {Seubert}}, \bibinfo {author} {\bibfnamefont {T.}~\bibnamefont {Bauer}}, \bibinfo {author} {\bibfnamefont {W.}~\bibnamefont {Rosenfeld}}, \bibinfo {author} {\bibfnamefont {W.}~\bibnamefont {Zhang}}, \bibinfo {author} {\bibfnamefont {C.}~\bibnamefont {Becher}},\ and\ \bibinfo {author} {\bibfnamefont {H.}~\bibnamefont {Weinfurter}},\ }\bibfield  {title} {\bibinfo {title} {Entangling single atoms over 33{\thinspace}km telecom fibre},\ }\href {https://doi.org/10.1038/s41586-022-04764-4} {\bibfield  {journal}
  {\bibinfo  {journal} {Nature}\ }\textbf {\bibinfo {volume} {607}},\ \bibinfo {pages} {69} (\bibinfo {year} {2022})}\BibitemShut {NoStop}%
\bibitem [{\citenamefont {Lei}\ \emph {et~al.}(2023)\citenamefont {Lei}, \citenamefont {Asadi}, \citenamefont {Zhong}, \citenamefont {Kuzmich}, \citenamefont {Simon},\ and\ \citenamefont {Hosseini}}]{Lei:2023}%
  \BibitemOpen
  \bibfield  {author} {\bibinfo {author} {\bibfnamefont {Y.}~\bibnamefont {Lei}}, \bibinfo {author} {\bibfnamefont {F.~K.}\ \bibnamefont {Asadi}}, \bibinfo {author} {\bibfnamefont {T.}~\bibnamefont {Zhong}}, \bibinfo {author} {\bibfnamefont {A.}~\bibnamefont {Kuzmich}}, \bibinfo {author} {\bibfnamefont {C.}~\bibnamefont {Simon}},\ and\ \bibinfo {author} {\bibfnamefont {M.}~\bibnamefont {Hosseini}},\ }\bibfield  {title} {\bibinfo {title} {Quantum optical memory for entanglement distribution},\ }\href {https://doi.org/10.1364/OPTICA.493732} {\bibfield  {journal} {\bibinfo  {journal} {Optica}\ }\textbf {\bibinfo {volume} {10}},\ \bibinfo {pages} {1511} (\bibinfo {year} {2023})}\BibitemShut {NoStop}%
\bibitem [{\citenamefont {Chen}\ \emph {et~al.}(2021)\citenamefont {Chen}, \citenamefont {Zhang}, \citenamefont {Chen}, \citenamefont {Cai}, \citenamefont {Liao}, \citenamefont {Zhang}, \citenamefont {Chen}, \citenamefont {Yin}, \citenamefont {Ren}, \citenamefont {Chen}, \citenamefont {Han}, \citenamefont {Yu}, \citenamefont {Liang}, \citenamefont {Zhou}, \citenamefont {Yuan}, \citenamefont {Zhao}, \citenamefont {Wang}, \citenamefont {Jiang}, \citenamefont {Zhang}, \citenamefont {Liu}, \citenamefont {Li}, \citenamefont {Shen}, \citenamefont {Cao}, \citenamefont {Lu}, \citenamefont {Shu}, \citenamefont {Wang}, \citenamefont {Li}, \citenamefont {Liu}, \citenamefont {Xu}, \citenamefont {Wang}, \citenamefont {Peng},\ and\ \citenamefont {Pan}}]{Chen:2021}%
  \BibitemOpen
  \bibfield  {author} {\bibinfo {author} {\bibfnamefont {Y.-A.}\ \bibnamefont {Chen}}, \bibinfo {author} {\bibfnamefont {Q.}~\bibnamefont {Zhang}}, \bibinfo {author} {\bibfnamefont {T.-Y.}\ \bibnamefont {Chen}}, \bibinfo {author} {\bibfnamefont {W.-Q.}\ \bibnamefont {Cai}}, \bibinfo {author} {\bibfnamefont {S.-K.}\ \bibnamefont {Liao}}, \bibinfo {author} {\bibfnamefont {J.}~\bibnamefont {Zhang}}, \bibinfo {author} {\bibfnamefont {K.}~\bibnamefont {Chen}}, \bibinfo {author} {\bibfnamefont {J.}~\bibnamefont {Yin}}, \bibinfo {author} {\bibfnamefont {J.-G.}\ \bibnamefont {Ren}}, \bibinfo {author} {\bibfnamefont {Z.}~\bibnamefont {Chen}}, \bibinfo {author} {\bibfnamefont {S.-L.}\ \bibnamefont {Han}}, \bibinfo {author} {\bibfnamefont {Q.}~\bibnamefont {Yu}}, \bibinfo {author} {\bibfnamefont {K.}~\bibnamefont {Liang}}, \bibinfo {author} {\bibfnamefont {F.}~\bibnamefont {Zhou}}, \bibinfo {author} {\bibfnamefont {X.}~\bibnamefont {Yuan}}, \bibinfo {author} {\bibfnamefont {M.-S.}\ \bibnamefont {Zhao}}, \bibinfo
  {author} {\bibfnamefont {T.-Y.}\ \bibnamefont {Wang}}, \bibinfo {author} {\bibfnamefont {X.}~\bibnamefont {Jiang}}, \bibinfo {author} {\bibfnamefont {L.}~\bibnamefont {Zhang}}, \bibinfo {author} {\bibfnamefont {W.-Y.}\ \bibnamefont {Liu}}, \bibinfo {author} {\bibfnamefont {Y.}~\bibnamefont {Li}}, \bibinfo {author} {\bibfnamefont {Q.}~\bibnamefont {Shen}}, \bibinfo {author} {\bibfnamefont {Y.}~\bibnamefont {Cao}}, \bibinfo {author} {\bibfnamefont {C.-Y.}\ \bibnamefont {Lu}}, \bibinfo {author} {\bibfnamefont {R.}~\bibnamefont {Shu}}, \bibinfo {author} {\bibfnamefont {J.-Y.}\ \bibnamefont {Wang}}, \bibinfo {author} {\bibfnamefont {L.}~\bibnamefont {Li}}, \bibinfo {author} {\bibfnamefont {N.-L.}\ \bibnamefont {Liu}}, \bibinfo {author} {\bibfnamefont {F.}~\bibnamefont {Xu}}, \bibinfo {author} {\bibfnamefont {X.-B.}\ \bibnamefont {Wang}}, \bibinfo {author} {\bibfnamefont {C.-Z.}\ \bibnamefont {Peng}},\ and\ \bibinfo {author} {\bibfnamefont {J.-W.}\ \bibnamefont {Pan}},\ }\bibfield  {title} {\bibinfo {title} {An
  integrated space-to-ground quantum communication network over 4,600 kilometres},\ }\href {https://doi.org/10.1038/s41586-020-03093-8} {\bibfield  {journal} {\bibinfo  {journal} {Nature}\ }\textbf {\bibinfo {volume} {589}},\ \bibinfo {pages} {214} (\bibinfo {year} {2021})}\BibitemShut {NoStop}%
\bibitem [{\citenamefont {Reim}\ \emph {et~al.}(2010)\citenamefont {Reim}, \citenamefont {Nunn}, \citenamefont {Lorenz}, \citenamefont {Sussman}, \citenamefont {Lee}, \citenamefont {Langford}, \citenamefont {Jaksch},\ and\ \citenamefont {Walmsley}}]{Reim:2010}%
  \BibitemOpen
  \bibfield  {author} {\bibinfo {author} {\bibfnamefont {K.~F.}\ \bibnamefont {Reim}}, \bibinfo {author} {\bibfnamefont {J.}~\bibnamefont {Nunn}}, \bibinfo {author} {\bibfnamefont {V.~O.}\ \bibnamefont {Lorenz}}, \bibinfo {author} {\bibfnamefont {B.~J.}\ \bibnamefont {Sussman}}, \bibinfo {author} {\bibfnamefont {K.~C.}\ \bibnamefont {Lee}}, \bibinfo {author} {\bibfnamefont {N.~K.}\ \bibnamefont {Langford}}, \bibinfo {author} {\bibfnamefont {D.}~\bibnamefont {Jaksch}},\ and\ \bibinfo {author} {\bibfnamefont {I.~A.}\ \bibnamefont {Walmsley}},\ }\bibfield  {title} {\bibinfo {title} {Towards high-speed optical quantum memories},\ }\href {https://doi.org/10.1038/nphoton.2010.30} {\bibfield  {journal} {\bibinfo  {journal} {Nature Photonics}\ }\textbf {\bibinfo {volume} {4}},\ \bibinfo {pages} {218} (\bibinfo {year} {2010})}\BibitemShut {NoStop}%
\bibitem [{\citenamefont {{Kai Shinbrough}}\ \emph {et~al.}(2023)\citenamefont {{Kai Shinbrough}}, \citenamefont {{Donny R. Pearson}}, \citenamefont {{Bin Fang}}, \citenamefont {{Elizabeth A. Goldschmidt}},\ and\ \citenamefont {{Virginia O. Lorenz}}}]{KaiShinbrough.2023}%
  \BibitemOpen
  \bibfield  {author} {\bibinfo {author} {\bibnamefont {{Kai Shinbrough}}}, \bibinfo {author} {\bibnamefont {{Donny R. Pearson}}}, \bibinfo {author} {\bibnamefont {{Bin Fang}}}, \bibinfo {author} {\bibnamefont {{Elizabeth A. Goldschmidt}}},\ and\ \bibinfo {author} {\bibnamefont {{Virginia O. Lorenz}}},\ }\bibfield  {title} {\bibinfo {title} {Chapter five - broadband quantum memory in atomic ensembles},\ }in\ \href {https://doi.org/10.1016/bs.aamop.2023.04.001} {\emph {\bibinfo {booktitle} {Advances in Atomic, Molecular, and Optical Physics}}},\ \bibinfo {series} {Advances In Atomic, Molecular, and Optical Physics}, Vol.~\bibinfo {volume} {72},\ \bibinfo {editor} {edited by\ \bibinfo {editor} {\bibnamefont {{Louis F. DiMauro}}}, \bibinfo {editor} {\bibnamefont {{H{\'e}l{\`e}ne Perrin}}},\ and\ \bibinfo {editor} {\bibnamefont {{Susanne F. Yelin}}}}\ (\bibinfo  {publisher} {{Academic Press}},\ \bibinfo {year} {2023})\ pp.\ \bibinfo {pages} {297--360}\BibitemShut {NoStop}%
\bibitem [{\citenamefont {Me{\ss}ner}\ \emph {et~al.}(2023)\citenamefont {Me{\ss}ner}, \citenamefont {Robertson}, \citenamefont {Esguerra}, \citenamefont {L\"{u}dge},\ and\ \citenamefont {Wolters}}]{MeSSner:23}%
  \BibitemOpen
  \bibfield  {author} {\bibinfo {author} {\bibfnamefont {L.}~\bibnamefont {Me{\ss}ner}}, \bibinfo {author} {\bibfnamefont {E.}~\bibnamefont {Robertson}}, \bibinfo {author} {\bibfnamefont {L.}~\bibnamefont {Esguerra}}, \bibinfo {author} {\bibfnamefont {K.}~\bibnamefont {L\"{u}dge}},\ and\ \bibinfo {author} {\bibfnamefont {J.}~\bibnamefont {Wolters}},\ }\bibfield  {title} {\bibinfo {title} {Multiplexed random-access optical memory in warm cesium vapor},\ }\href {https://doi.org/10.1364/OE.483642} {\bibfield  {journal} {\bibinfo  {journal} {Opt. Express}\ }\textbf {\bibinfo {volume} {31}},\ \bibinfo {pages} {10150} (\bibinfo {year} {2023})}\BibitemShut {NoStop}%
\bibitem [{\citenamefont {Wang}\ \emph {et~al.}(2022)\citenamefont {Wang}, \citenamefont {Craddock}, \citenamefont {Sekelsky}, \citenamefont {Flament},\ and\ \citenamefont {Namazi}}]{Wang2022}%
  \BibitemOpen
  \bibfield  {author} {\bibinfo {author} {\bibfnamefont {Y.}~\bibnamefont {Wang}}, \bibinfo {author} {\bibfnamefont {A.~N.}\ \bibnamefont {Craddock}}, \bibinfo {author} {\bibfnamefont {R.}~\bibnamefont {Sekelsky}}, \bibinfo {author} {\bibfnamefont {M.}~\bibnamefont {Flament}},\ and\ \bibinfo {author} {\bibfnamefont {M.}~\bibnamefont {Namazi}},\ }\bibfield  {title} {\bibinfo {title} {Field-deployable quantum memory for quantum networking},\ }\href {https://doi.org/10.1103/PhysRevApplied.18.044058} {\bibfield  {journal} {\bibinfo  {journal} {Phys. Rev. Appl.}\ }\textbf {\bibinfo {volume} {18}},\ \bibinfo {pages} {044058} (\bibinfo {year} {2022})}\BibitemShut {NoStop}%
\bibitem [{\citenamefont {Wallnöfer}\ \emph {et~al.}(2022)\citenamefont {Wallnöfer}, \citenamefont {Hahn}, \citenamefont {Gündoğan}, \citenamefont {Sidhu}, \citenamefont {Wiesner}, \citenamefont {Walk}, \citenamefont {Eisert},\ and\ \citenamefont {Wolters}}]{Wallnoefer:2022}%
  \BibitemOpen
  \bibfield  {author} {\bibinfo {author} {\bibfnamefont {J.}~\bibnamefont {Wallnöfer}}, \bibinfo {author} {\bibfnamefont {F.}~\bibnamefont {Hahn}}, \bibinfo {author} {\bibfnamefont {M.}~\bibnamefont {Gündoğan}}, \bibinfo {author} {\bibfnamefont {J.~S.}\ \bibnamefont {Sidhu}}, \bibinfo {author} {\bibfnamefont {F.}~\bibnamefont {Wiesner}}, \bibinfo {author} {\bibfnamefont {N.}~\bibnamefont {Walk}}, \bibinfo {author} {\bibfnamefont {J.}~\bibnamefont {Eisert}},\ and\ \bibinfo {author} {\bibfnamefont {J.}~\bibnamefont {Wolters}},\ }\bibfield  {title} {\bibinfo {title} {Simulating quantum repeater strategies for multiple satellites},\ }\href {https://doi.org/10.1038/s42005-022-00945-9} {\bibfield  {journal} {\bibinfo  {journal} {Communications Physics}\ }\textbf {\bibinfo {volume} {5}},\ \bibinfo {pages} {169} (\bibinfo {year} {2022})}\BibitemShut {NoStop}%
\bibitem [{\citenamefont {G{\"u}ndo{\u{g}}an}\ \emph {et~al.}(2021)\citenamefont {G{\"u}ndo{\u{g}}an}, \citenamefont {Sidhu}, \citenamefont {Henderson}, \citenamefont {Mazzarella}, \citenamefont {Wolters}, \citenamefont {Oi},\ and\ \citenamefont {Krutzik}}]{Guendogan:2021}%
  \BibitemOpen
  \bibfield  {author} {\bibinfo {author} {\bibfnamefont {M.}~\bibnamefont {G{\"u}ndo{\u{g}}an}}, \bibinfo {author} {\bibfnamefont {J.~S.}\ \bibnamefont {Sidhu}}, \bibinfo {author} {\bibfnamefont {V.}~\bibnamefont {Henderson}}, \bibinfo {author} {\bibfnamefont {L.}~\bibnamefont {Mazzarella}}, \bibinfo {author} {\bibfnamefont {J.}~\bibnamefont {Wolters}}, \bibinfo {author} {\bibfnamefont {D.~K.~L.}\ \bibnamefont {Oi}},\ and\ \bibinfo {author} {\bibfnamefont {M.}~\bibnamefont {Krutzik}},\ }\bibfield  {title} {\bibinfo {title} {Proposal for space-borne quantum memories for global quantum networking},\ }\href {https://doi.org/10.1038/s41534-021-00460-9} {\bibfield  {journal} {\bibinfo  {journal} {npj Quantum Information}\ }\textbf {\bibinfo {volume} {7}},\ \bibinfo {pages} {128} (\bibinfo {year} {2021})}\BibitemShut {NoStop}%
\bibitem [{\citenamefont {Katz}\ and\ \citenamefont {Firstenberg}(2018)}]{Katz.2018}%
  \BibitemOpen
  \bibfield  {author} {\bibinfo {author} {\bibfnamefont {O.}~\bibnamefont {Katz}}\ and\ \bibinfo {author} {\bibfnamefont {O.}~\bibnamefont {Firstenberg}},\ }\bibfield  {title} {\bibinfo {title} {Light storage for one second in room-temperature alkali vapor},\ }\href {https://doi.org/10.1038/s41467-018-04458-4} {\bibfield  {journal} {\bibinfo  {journal} {Nature Communications}\ }\textbf {\bibinfo {volume} {9}},\ \bibinfo {pages} {2074} (\bibinfo {year} {2018})}\BibitemShut {NoStop}%
\bibitem [{\citenamefont {Buser}\ \emph {et~al.}(2022)\citenamefont {Buser}, \citenamefont {Mottola}, \citenamefont {Cotting}, \citenamefont {Wolters},\ and\ \citenamefont {Treutlein}}]{Buser:2023}%
  \BibitemOpen
  \bibfield  {author} {\bibinfo {author} {\bibfnamefont {G.}~\bibnamefont {Buser}}, \bibinfo {author} {\bibfnamefont {R.}~\bibnamefont {Mottola}}, \bibinfo {author} {\bibfnamefont {B.}~\bibnamefont {Cotting}}, \bibinfo {author} {\bibfnamefont {J.}~\bibnamefont {Wolters}},\ and\ \bibinfo {author} {\bibfnamefont {P.}~\bibnamefont {Treutlein}},\ }\bibfield  {title} {\bibinfo {title} {Single-photon storage in a ground-state vapor cell quantum memory},\ }\href {https://doi.org/10.1103/PRXQuantum.3.020349} {\bibfield  {journal} {\bibinfo  {journal} {PRX Quantum}\ }\textbf {\bibinfo {volume} {3}},\ \bibinfo {pages} {020349} (\bibinfo {year} {2022})}\BibitemShut {NoStop}%
\bibitem [{\citenamefont {Dideriksen}\ \emph {et~al.}(2021)\citenamefont {Dideriksen}, \citenamefont {Schmieg}, \citenamefont {Zugenmaier},\ and\ \citenamefont {Polzik}}]{Dideriksen.2021}%
  \BibitemOpen
  \bibfield  {author} {\bibinfo {author} {\bibfnamefont {K.~B.}\ \bibnamefont {Dideriksen}}, \bibinfo {author} {\bibfnamefont {R.}~\bibnamefont {Schmieg}}, \bibinfo {author} {\bibfnamefont {M.}~\bibnamefont {Zugenmaier}},\ and\ \bibinfo {author} {\bibfnamefont {E.~S.}\ \bibnamefont {Polzik}},\ }\bibfield  {title} {\bibinfo {title} {Room-temperature single-photon source with near-millisecond built-in memory},\ }\href {https://doi.org/10.1038/s41467-021-24033-8} {\bibfield  {journal} {\bibinfo  {journal} {Nature Communications}\ }\textbf {\bibinfo {volume} {12}},\ \bibinfo {pages} {3699} (\bibinfo {year} {2021})}\BibitemShut {NoStop}%
\bibitem [{\citenamefont {Davidson}\ \emph {et~al.}(2023{\natexlab{b}})\citenamefont {Davidson}, \citenamefont {Yogev}, \citenamefont {Poem},\ and\ \citenamefont {Firstenberg}}]{Davidson:2023-2}%
  \BibitemOpen
  \bibfield  {author} {\bibinfo {author} {\bibfnamefont {O.}~\bibnamefont {Davidson}}, \bibinfo {author} {\bibfnamefont {O.}~\bibnamefont {Yogev}}, \bibinfo {author} {\bibfnamefont {E.}~\bibnamefont {Poem}},\ and\ \bibinfo {author} {\bibfnamefont {O.}~\bibnamefont {Firstenberg}},\ }\href {https://doi.org/10.1038/s42005-023-01247-4} {\bibinfo {title} {Fast, noise-free atomic optical memory with 35-percent end-to-end efficiency}} (\bibinfo {year} {2023}{\natexlab{b}})\BibitemShut {NoStop}%
\bibitem [{\citenamefont {Kaczmarek}\ \emph {et~al.}(2018)\citenamefont {Kaczmarek}, \citenamefont {Ledingham}, \citenamefont {Brecht}, \citenamefont {Thomas}, \citenamefont {Thekkadath}, \citenamefont {Lazo-Arjona}, \citenamefont {Munns}, \citenamefont {Poem}, \citenamefont {Feizpour}, \citenamefont {Saunders}, \citenamefont {Nunn},\ and\ \citenamefont {Walmsley}}]{Kaczmarek:2018}%
  \BibitemOpen
  \bibfield  {author} {\bibinfo {author} {\bibfnamefont {K.~T.}\ \bibnamefont {Kaczmarek}}, \bibinfo {author} {\bibfnamefont {P.~M.}\ \bibnamefont {Ledingham}}, \bibinfo {author} {\bibfnamefont {B.}~\bibnamefont {Brecht}}, \bibinfo {author} {\bibfnamefont {S.~E.}\ \bibnamefont {Thomas}}, \bibinfo {author} {\bibfnamefont {G.~S.}\ \bibnamefont {Thekkadath}}, \bibinfo {author} {\bibfnamefont {O.}~\bibnamefont {Lazo-Arjona}}, \bibinfo {author} {\bibfnamefont {J.~H.~D.}\ \bibnamefont {Munns}}, \bibinfo {author} {\bibfnamefont {E.}~\bibnamefont {Poem}}, \bibinfo {author} {\bibfnamefont {A.}~\bibnamefont {Feizpour}}, \bibinfo {author} {\bibfnamefont {D.~J.}\ \bibnamefont {Saunders}}, \bibinfo {author} {\bibfnamefont {J.}~\bibnamefont {Nunn}},\ and\ \bibinfo {author} {\bibfnamefont {I.~A.}\ \bibnamefont {Walmsley}},\ }\bibfield  {title} {\bibinfo {title} {High-speed noise-free optical quantum memory},\ }\href {https://doi.org/10.1103/PhysRevA.97.042316} {\bibfield  {journal} {\bibinfo  {journal} {Phys. Rev. A}\
  }\textbf {\bibinfo {volume} {97}},\ \bibinfo {pages} {042316} (\bibinfo {year} {2018})}\BibitemShut {NoStop}%
\bibitem [{\citenamefont {Finkelstein}\ \emph {et~al.}(2018)\citenamefont {Finkelstein}, \citenamefont {Poem}, \citenamefont {Michel}, \citenamefont {Lahad},\ and\ \citenamefont {Firstenberg}}]{Finkelstein:2018}%
  \BibitemOpen
  \bibfield  {author} {\bibinfo {author} {\bibfnamefont {R.}~\bibnamefont {Finkelstein}}, \bibinfo {author} {\bibfnamefont {E.}~\bibnamefont {Poem}}, \bibinfo {author} {\bibfnamefont {O.}~\bibnamefont {Michel}}, \bibinfo {author} {\bibfnamefont {O.}~\bibnamefont {Lahad}},\ and\ \bibinfo {author} {\bibfnamefont {O.}~\bibnamefont {Firstenberg}},\ }\bibfield  {title} {\bibinfo {title} {Fast, noise-free memory for photon synchronization at room temperature},\ }\href {https://doi.org/10.1126/sciadv.aap8598} {\bibfield  {journal} {\bibinfo  {journal} {Science Advances}\ }\textbf {\bibinfo {volume} {4}},\ \bibinfo {pages} {eaap8598} (\bibinfo {year} {2018})}\BibitemShut {NoStop}%
\bibitem [{\citenamefont {Thomas}\ \emph {et~al.}(2023)\citenamefont {Thomas}, \citenamefont {Sagona-Stophel}, \citenamefont {Schofield}, \citenamefont {Walmsley},\ and\ \citenamefont {Ledingham}}]{Thomas:2023}%
  \BibitemOpen
  \bibfield  {author} {\bibinfo {author} {\bibfnamefont {S.}~\bibnamefont {Thomas}}, \bibinfo {author} {\bibfnamefont {S.}~\bibnamefont {Sagona-Stophel}}, \bibinfo {author} {\bibfnamefont {Z.}~\bibnamefont {Schofield}}, \bibinfo {author} {\bibfnamefont {I.}~\bibnamefont {Walmsley}},\ and\ \bibinfo {author} {\bibfnamefont {P.}~\bibnamefont {Ledingham}},\ }\bibfield  {title} {\bibinfo {title} {{S}ingle-{P}hoton-{C}ompatible {T}elecommunications-{B}and {Q}uantum {M}emory in a {H}ot {A}tomic {G}as},\ }\href {https://doi.org/10.1103/PhysRevApplied.19.L031005} {\bibfield  {journal} {\bibinfo  {journal} {Phys. Rev. Appl.}\ }\textbf {\bibinfo {volume} {19}},\ \bibinfo {pages} {L031005} (\bibinfo {year} {2023})}\BibitemShut {NoStop}%
\bibitem [{\citenamefont {Gschrey}\ \emph {et~al.}(2015)\citenamefont {Gschrey}, \citenamefont {Thoma}, \citenamefont {Schnauber}, \citenamefont {Seifried}, \citenamefont {Schmidt}, \citenamefont {Wohlfeil}, \citenamefont {Kr{\"u}ger}, \citenamefont {Schulze}, \citenamefont {Heindel}, \citenamefont {Burger}, \citenamefont {Schmidt}, \citenamefont {Strittmatter}, \citenamefont {Rodt},\ and\ \citenamefont {Reitzenstein}}]{Gschrey.2015}%
  \BibitemOpen
  \bibfield  {author} {\bibinfo {author} {\bibfnamefont {M.}~\bibnamefont {Gschrey}}, \bibinfo {author} {\bibfnamefont {A.}~\bibnamefont {Thoma}}, \bibinfo {author} {\bibfnamefont {P.}~\bibnamefont {Schnauber}}, \bibinfo {author} {\bibfnamefont {M.}~\bibnamefont {Seifried}}, \bibinfo {author} {\bibfnamefont {R.}~\bibnamefont {Schmidt}}, \bibinfo {author} {\bibfnamefont {B.}~\bibnamefont {Wohlfeil}}, \bibinfo {author} {\bibfnamefont {L.}~\bibnamefont {Kr{\"u}ger}}, \bibinfo {author} {\bibfnamefont {J.-H.}\ \bibnamefont {Schulze}}, \bibinfo {author} {\bibfnamefont {T.}~\bibnamefont {Heindel}}, \bibinfo {author} {\bibfnamefont {S.}~\bibnamefont {Burger}}, \bibinfo {author} {\bibfnamefont {F.}~\bibnamefont {Schmidt}}, \bibinfo {author} {\bibfnamefont {A.}~\bibnamefont {Strittmatter}}, \bibinfo {author} {\bibfnamefont {S.}~\bibnamefont {Rodt}},\ and\ \bibinfo {author} {\bibfnamefont {S.}~\bibnamefont {Reitzenstein}},\ }\bibfield  {title} {\bibinfo {title} {Highly indistinguishable photons from deterministic
  quantum-dot microlenses utilizing three-dimensional in situ electron-beam lithography},\ }\href {https://doi.org/10.1038/ncomms8662} {\bibfield  {journal} {\bibinfo  {journal} {Nature Communications}\ }\textbf {\bibinfo {volume} {6}},\ \bibinfo {pages} {7662} (\bibinfo {year} {2015})}\BibitemShut {NoStop}%
\bibitem [{\citenamefont {Heindel}\ \emph {et~al.}(2023)\citenamefont {Heindel}, \citenamefont {Kim}, \citenamefont {Gregersen}, \citenamefont {Rastelli},\ and\ \citenamefont {Reitzenstein}}]{Heindel:2023}%
  \BibitemOpen
  \bibfield  {author} {\bibinfo {author} {\bibfnamefont {T.}~\bibnamefont {Heindel}}, \bibinfo {author} {\bibfnamefont {J.-H.}\ \bibnamefont {Kim}}, \bibinfo {author} {\bibfnamefont {N.}~\bibnamefont {Gregersen}}, \bibinfo {author} {\bibfnamefont {A.}~\bibnamefont {Rastelli}},\ and\ \bibinfo {author} {\bibfnamefont {S.}~\bibnamefont {Reitzenstein}},\ }\bibfield  {title} {\bibinfo {title} {Quantum dots for photonic quantum information technology},\ }\href {https://doi.org/10.1364/AOP.490091} {\bibfield  {journal} {\bibinfo  {journal} {Adv. Opt. Photon.}\ }\textbf {\bibinfo {volume} {15}},\ \bibinfo {pages} {613} (\bibinfo {year} {2023})}\BibitemShut {NoStop}%
\bibitem [{\citenamefont {Uppu}\ \emph {et~al.}(2021)\citenamefont {Uppu}, \citenamefont {Midolo}, \citenamefont {Zhou}, \citenamefont {Carolan},\ and\ \citenamefont {Lodahl}}]{Uppu:2021}%
  \BibitemOpen
  \bibfield  {author} {\bibinfo {author} {\bibfnamefont {R.}~\bibnamefont {Uppu}}, \bibinfo {author} {\bibfnamefont {L.}~\bibnamefont {Midolo}}, \bibinfo {author} {\bibfnamefont {X.}~\bibnamefont {Zhou}}, \bibinfo {author} {\bibfnamefont {J.}~\bibnamefont {Carolan}},\ and\ \bibinfo {author} {\bibfnamefont {P.}~\bibnamefont {Lodahl}},\ }\bibfield  {title} {\bibinfo {title} {Quantum-dot-based deterministic photon--emitter interfaces for scalable photonic quantum technology},\ }\href {https://doi.org/10.1038/s41565-021-00965-6} {\bibfield  {journal} {\bibinfo  {journal} {Nature Nanotechnology}\ }\textbf {\bibinfo {volume} {16}},\ \bibinfo {pages} {1308} (\bibinfo {year} {2021})}\BibitemShut {NoStop}%
\bibitem [{\citenamefont {Wang}\ \emph {et~al.}(2019)\citenamefont {Wang}, \citenamefont {He}, \citenamefont {Chung}, \citenamefont {Hu}, \citenamefont {Yu}, \citenamefont {Chen}, \citenamefont {Ding}, \citenamefont {Chen}, \citenamefont {Qin}, \citenamefont {Yang}, \citenamefont {Liu}, \citenamefont {Duan}, \citenamefont {Li}, \citenamefont {Gerhardt}, \citenamefont {Winkler}, \citenamefont {Jurkat}, \citenamefont {Wang}, \citenamefont {Gregersen}, \citenamefont {Huo}, \citenamefont {Dai}, \citenamefont {Yu}, \citenamefont {H{\"o}fling}, \citenamefont {Lu},\ and\ \citenamefont {Pan}}]{Wang.2019}%
  \BibitemOpen
  \bibfield  {author} {\bibinfo {author} {\bibfnamefont {H.}~\bibnamefont {Wang}}, \bibinfo {author} {\bibfnamefont {Y.-M.}\ \bibnamefont {He}}, \bibinfo {author} {\bibfnamefont {T.-H.}\ \bibnamefont {Chung}}, \bibinfo {author} {\bibfnamefont {H.}~\bibnamefont {Hu}}, \bibinfo {author} {\bibfnamefont {Y.}~\bibnamefont {Yu}}, \bibinfo {author} {\bibfnamefont {S.}~\bibnamefont {Chen}}, \bibinfo {author} {\bibfnamefont {X.}~\bibnamefont {Ding}}, \bibinfo {author} {\bibfnamefont {M.-C.}\ \bibnamefont {Chen}}, \bibinfo {author} {\bibfnamefont {J.}~\bibnamefont {Qin}}, \bibinfo {author} {\bibfnamefont {X.}~\bibnamefont {Yang}}, \bibinfo {author} {\bibfnamefont {R.-Z.}\ \bibnamefont {Liu}}, \bibinfo {author} {\bibfnamefont {Z.-C.}\ \bibnamefont {Duan}}, \bibinfo {author} {\bibfnamefont {J.-P.}\ \bibnamefont {Li}}, \bibinfo {author} {\bibfnamefont {S.}~\bibnamefont {Gerhardt}}, \bibinfo {author} {\bibfnamefont {K.}~\bibnamefont {Winkler}}, \bibinfo {author} {\bibfnamefont {J.}~\bibnamefont {Jurkat}}, \bibinfo {author}
  {\bibfnamefont {L.-J.}\ \bibnamefont {Wang}}, \bibinfo {author} {\bibfnamefont {N.}~\bibnamefont {Gregersen}}, \bibinfo {author} {\bibfnamefont {Y.-H.}\ \bibnamefont {Huo}}, \bibinfo {author} {\bibfnamefont {Q.}~\bibnamefont {Dai}}, \bibinfo {author} {\bibfnamefont {S.}~\bibnamefont {Yu}}, \bibinfo {author} {\bibfnamefont {S.}~\bibnamefont {H{\"o}fling}}, \bibinfo {author} {\bibfnamefont {C.-Y.}\ \bibnamefont {Lu}},\ and\ \bibinfo {author} {\bibfnamefont {J.-W.}\ \bibnamefont {Pan}},\ }\bibfield  {title} {\bibinfo {title} {Towards optimal single-photon sources from polarized microcavities},\ }\href {https://doi.org/10.1038/s41566-019-0494-3} {\bibfield  {journal} {\bibinfo  {journal} {Nature Photonics}\ }\textbf {\bibinfo {volume} {13}},\ \bibinfo {pages} {770} (\bibinfo {year} {2019})}\BibitemShut {NoStop}%
\bibitem [{\citenamefont {Nawrath}\ \emph {et~al.}(2023)\citenamefont {Nawrath}, \citenamefont {Joos}, \citenamefont {Kolatschek}, \citenamefont {Bauer}, \citenamefont {Pruy}, \citenamefont {Hornung}, \citenamefont {Fischer}, \citenamefont {Huang}, \citenamefont {Vijayan}, \citenamefont {Sittig}, \citenamefont {Jetter}, \citenamefont {Portalupi},\ and\ \citenamefont {Michler}}]{Nawrath.2023}%
  \BibitemOpen
  \bibfield  {author} {\bibinfo {author} {\bibfnamefont {C.}~\bibnamefont {Nawrath}}, \bibinfo {author} {\bibfnamefont {R.}~\bibnamefont {Joos}}, \bibinfo {author} {\bibfnamefont {S.}~\bibnamefont {Kolatschek}}, \bibinfo {author} {\bibfnamefont {S.}~\bibnamefont {Bauer}}, \bibinfo {author} {\bibfnamefont {P.}~\bibnamefont {Pruy}}, \bibinfo {author} {\bibfnamefont {F.}~\bibnamefont {Hornung}}, \bibinfo {author} {\bibfnamefont {J.}~\bibnamefont {Fischer}}, \bibinfo {author} {\bibfnamefont {J.}~\bibnamefont {Huang}}, \bibinfo {author} {\bibfnamefont {P.}~\bibnamefont {Vijayan}}, \bibinfo {author} {\bibfnamefont {R.}~\bibnamefont {Sittig}}, \bibinfo {author} {\bibfnamefont {M.}~\bibnamefont {Jetter}}, \bibinfo {author} {\bibfnamefont {S.~L.}\ \bibnamefont {Portalupi}},\ and\ \bibinfo {author} {\bibfnamefont {P.}~\bibnamefont {Michler}},\ }\bibfield  {title} {\bibinfo {title} {Bright source of purcell-enhanced, triggered, single photons in the telecom c-band},\ }\href {https://doi.org/10.1002/qute.202300111}
  {\bibfield  {journal} {\bibinfo  {journal} {Advanced Quantum Technologies}\ }\textbf {\bibinfo {volume} {6}},\ \bibinfo {pages} {2300111} (\bibinfo {year} {2023})}\BibitemShut {NoStop}%
\bibitem [{\citenamefont {Tomm}\ \emph {et~al.}(2021)\citenamefont {Tomm}, \citenamefont {Javadi}, \citenamefont {Antoniadis}, \citenamefont {Najer}, \citenamefont {L{\"o}bl}, \citenamefont {Korsch}, \citenamefont {Schott}, \citenamefont {Valentin}, \citenamefont {Wieck}, \citenamefont {Ludwig},\ and\ \citenamefont {Warburton}}]{Tomm.2021}%
  \BibitemOpen
  \bibfield  {author} {\bibinfo {author} {\bibfnamefont {N.}~\bibnamefont {Tomm}}, \bibinfo {author} {\bibfnamefont {A.}~\bibnamefont {Javadi}}, \bibinfo {author} {\bibfnamefont {N.~O.}\ \bibnamefont {Antoniadis}}, \bibinfo {author} {\bibfnamefont {D.}~\bibnamefont {Najer}}, \bibinfo {author} {\bibfnamefont {M.~C.}\ \bibnamefont {L{\"o}bl}}, \bibinfo {author} {\bibfnamefont {A.~R.}\ \bibnamefont {Korsch}}, \bibinfo {author} {\bibfnamefont {R.}~\bibnamefont {Schott}}, \bibinfo {author} {\bibfnamefont {S.~R.}\ \bibnamefont {Valentin}}, \bibinfo {author} {\bibfnamefont {A.~D.}\ \bibnamefont {Wieck}}, \bibinfo {author} {\bibfnamefont {A.}~\bibnamefont {Ludwig}},\ and\ \bibinfo {author} {\bibfnamefont {R.~J.}\ \bibnamefont {Warburton}},\ }\bibfield  {title} {\bibinfo {title} {A bright and fast source of coherent single photons},\ }\href {https://doi.org/10.1038/s41565-020-00831-x} {\bibfield  {journal} {\bibinfo  {journal} {Nature Nanotechnology}\ }\textbf {\bibinfo {volume} {16}},\ \bibinfo {pages} {399}
  (\bibinfo {year} {2021})}\BibitemShut {NoStop}%
\bibitem [{\citenamefont {Wildmann}\ \emph {et~al.}(2015)\citenamefont {Wildmann}, \citenamefont {Trotta}, \citenamefont {Mart\'{\i}n-S\'anchez}, \citenamefont {Zallo}, \citenamefont {O'Steen}, \citenamefont {Schmidt},\ and\ \citenamefont {Rastelli}}]{Wildmann:2015}%
  \BibitemOpen
  \bibfield  {author} {\bibinfo {author} {\bibfnamefont {J.~S.}\ \bibnamefont {Wildmann}}, \bibinfo {author} {\bibfnamefont {R.}~\bibnamefont {Trotta}}, \bibinfo {author} {\bibfnamefont {J.}~\bibnamefont {Mart\'{\i}n-S\'anchez}}, \bibinfo {author} {\bibfnamefont {E.}~\bibnamefont {Zallo}}, \bibinfo {author} {\bibfnamefont {M.}~\bibnamefont {O'Steen}}, \bibinfo {author} {\bibfnamefont {O.~G.}\ \bibnamefont {Schmidt}},\ and\ \bibinfo {author} {\bibfnamefont {A.}~\bibnamefont {Rastelli}},\ }\bibfield  {title} {\bibinfo {title} {{Atomic clouds as spectrally selective and tunable delay lines for single photons from quantum dots}},\ }\href {https://doi.org/10.1103/PhysRevB.92.235306} {\bibfield  {journal} {\bibinfo  {journal} {Phys. Rev. B}\ }\textbf {\bibinfo {volume} {92}},\ \bibinfo {pages} {235306} (\bibinfo {year} {2015})}\BibitemShut {NoStop}%
\bibitem [{\citenamefont {Bremer}\ \emph {et~al.}(2020)\citenamefont {Bremer}, \citenamefont {Fischbach}, \citenamefont {Park}, \citenamefont {Rodt}, \citenamefont {Song}, \citenamefont {Heindel},\ and\ \citenamefont {Reitzenstein}}]{Bremer:2020}%
  \BibitemOpen
  \bibfield  {author} {\bibinfo {author} {\bibfnamefont {L.}~\bibnamefont {Bremer}}, \bibinfo {author} {\bibfnamefont {S.}~\bibnamefont {Fischbach}}, \bibinfo {author} {\bibfnamefont {S.-I.}\ \bibnamefont {Park}}, \bibinfo {author} {\bibfnamefont {S.}~\bibnamefont {Rodt}}, \bibinfo {author} {\bibfnamefont {J.-D.}\ \bibnamefont {Song}}, \bibinfo {author} {\bibfnamefont {T.}~\bibnamefont {Heindel}},\ and\ \bibinfo {author} {\bibfnamefont {S.}~\bibnamefont {Reitzenstein}},\ }\bibfield  {title} {\bibinfo {title} {{Cesium-Vapor-Based Delay of Single Photons Emitted by Deterministically Fabricated Quantum Dot Microlenses}},\ }\href {https://doi.org/https://doi.org/10.1002/qute.201900071} {\bibfield  {journal} {\bibinfo  {journal} {Advanced Quantum Technologies}\ }\textbf {\bibinfo {volume} {3}},\ \bibinfo {pages} {1900071} (\bibinfo {year} {2020})}\BibitemShut {NoStop}%
\bibitem [{\citenamefont {Vural}\ \emph {et~al.}(2021)\citenamefont {Vural}, \citenamefont {Seyfferle}, \citenamefont {Gerhardt}, \citenamefont {Jetter}, \citenamefont {Portalupi},\ and\ \citenamefont {Michler}}]{Vural:2021}%
  \BibitemOpen
  \bibfield  {author} {\bibinfo {author} {\bibfnamefont {H.}~\bibnamefont {Vural}}, \bibinfo {author} {\bibfnamefont {S.}~\bibnamefont {Seyfferle}}, \bibinfo {author} {\bibfnamefont {I.}~\bibnamefont {Gerhardt}}, \bibinfo {author} {\bibfnamefont {M.}~\bibnamefont {Jetter}}, \bibinfo {author} {\bibfnamefont {S.~L.}\ \bibnamefont {Portalupi}},\ and\ \bibinfo {author} {\bibfnamefont {P.}~\bibnamefont {Michler}},\ }\bibfield  {title} {\bibinfo {title} {Delaying two-photon fock states in hot cesium vapor using single photons generated on demand from a semiconductor quantum dot},\ }\href {https://doi.org/10.1103/PhysRevB.103.195304} {\bibfield  {journal} {\bibinfo  {journal} {Phys. Rev. B}\ }\textbf {\bibinfo {volume} {103}},\ \bibinfo {pages} {195304} (\bibinfo {year} {2021})}\BibitemShut {NoStop}%
\bibitem [{\citenamefont {Kroh}\ \emph {et~al.}(2019)\citenamefont {Kroh}, \citenamefont {Wolters}, \citenamefont {Ahlrichs}, \citenamefont {Schell}, \citenamefont {Thoma}, \citenamefont {Reitzenstein}, \citenamefont {Wildmann}, \citenamefont {Zallo}, \citenamefont {Trotta}, \citenamefont {Rastelli}, \citenamefont {Schmidt},\ and\ \citenamefont {Benson}}]{Kroh:2019}%
  \BibitemOpen
  \bibfield  {author} {\bibinfo {author} {\bibfnamefont {T.}~\bibnamefont {Kroh}}, \bibinfo {author} {\bibfnamefont {J.}~\bibnamefont {Wolters}}, \bibinfo {author} {\bibfnamefont {A.}~\bibnamefont {Ahlrichs}}, \bibinfo {author} {\bibfnamefont {A.~W.}\ \bibnamefont {Schell}}, \bibinfo {author} {\bibfnamefont {A.}~\bibnamefont {Thoma}}, \bibinfo {author} {\bibfnamefont {S.}~\bibnamefont {Reitzenstein}}, \bibinfo {author} {\bibfnamefont {J.~S.}\ \bibnamefont {Wildmann}}, \bibinfo {author} {\bibfnamefont {E.}~\bibnamefont {Zallo}}, \bibinfo {author} {\bibfnamefont {R.}~\bibnamefont {Trotta}}, \bibinfo {author} {\bibfnamefont {A.}~\bibnamefont {Rastelli}}, \bibinfo {author} {\bibfnamefont {O.~G.}\ \bibnamefont {Schmidt}},\ and\ \bibinfo {author} {\bibfnamefont {O.}~\bibnamefont {Benson}},\ }\bibfield  {title} {\bibinfo {title} {{Slow and fast single photons from a quantum dot interacting with the excited state hyperfine structure of the Cesium D1-line}},\ }\href {https://doi.org/10.1038/s41598-019-50062-x}
  {\bibfield  {journal} {\bibinfo  {journal} {Scientific Reports}\ }\textbf {\bibinfo {volume} {9}},\ \bibinfo {pages} {13728} (\bibinfo {year} {2019})}\BibitemShut {NoStop}%
\bibitem [{\citenamefont {{Cui}}\ \emph {et~al.}(2023)\citenamefont {{Cui}}, \citenamefont {{Schweickert}}, \citenamefont {{J{\"o}ns}}, \citenamefont {{Namazi}}, \citenamefont {{Lettner}}, \citenamefont {{Zeuner}}, \citenamefont {{Scavuzzo Monta{\~n}a}}, \citenamefont {{Filipe Covre da Silva}}, \citenamefont {{Reindl}}, \citenamefont {{Huang}}, \citenamefont {{Trotta}}, \citenamefont {{Rastelli}}, \citenamefont {{Zwiller}},\ and\ \citenamefont {{Figueroa}}}]{Cui:2023}%
  \BibitemOpen
  \bibfield  {author} {\bibinfo {author} {\bibfnamefont {G.-D.}\ \bibnamefont {{Cui}}}, \bibinfo {author} {\bibfnamefont {L.}~\bibnamefont {{Schweickert}}}, \bibinfo {author} {\bibfnamefont {K.~D.}\ \bibnamefont {{J{\"o}ns}}}, \bibinfo {author} {\bibfnamefont {M.}~\bibnamefont {{Namazi}}}, \bibinfo {author} {\bibfnamefont {T.}~\bibnamefont {{Lettner}}}, \bibinfo {author} {\bibfnamefont {K.~D.}\ \bibnamefont {{Zeuner}}}, \bibinfo {author} {\bibfnamefont {L.}~\bibnamefont {{Scavuzzo Monta{\~n}a}}}, \bibinfo {author} {\bibfnamefont {S.}~\bibnamefont {{Filipe Covre da Silva}}}, \bibinfo {author} {\bibfnamefont {M.}~\bibnamefont {{Reindl}}}, \bibinfo {author} {\bibfnamefont {H.}~\bibnamefont {{Huang}}}, \bibinfo {author} {\bibfnamefont {R.}~\bibnamefont {{Trotta}}}, \bibinfo {author} {\bibfnamefont {A.}~\bibnamefont {{Rastelli}}}, \bibinfo {author} {\bibfnamefont {V.}~\bibnamefont {{Zwiller}}},\ and\ \bibinfo {author} {\bibfnamefont {E.}~\bibnamefont {{Figueroa}}},\ }\bibfield  {title} {\bibinfo {title} {{Coherent
  Quantum Interconnection between On-Demand Quantum Dot Single Photons and a Resonant Atomic Quantum Memory}},\ }\href {https://doi.org/10.48550/arXiv.2301.10326} {\bibfield  {journal} {\bibinfo  {journal} {arXiv e-prints}\ ,\ \bibinfo {eid} {arXiv:2301.10326}} (\bibinfo {year} {2023})},\ \Eprint {https://arxiv.org/abs/2301.10326} {arXiv:2301.10326 [quant-ph]} \BibitemShut {NoStop}%
\bibitem [{\citenamefont {{Al Maruf}}\ \emph {et~al.}(2023)\citenamefont {{Al Maruf}}, \citenamefont {{Venuturumilli}}, \citenamefont {{Bharadwaj}}, \citenamefont {{Anderson}}, \citenamefont {{Qiu}}, \citenamefont {{Yuan}}, \citenamefont {{Zeeshan}}, \citenamefont {{Semnani}}, \citenamefont {{Poole}}, \citenamefont {{Dalacu}}, \citenamefont {{Resch}}, \citenamefont {{Reimer}},\ and\ \citenamefont {{Bajcsy}}}]{AlMaruf:2023}%
  \BibitemOpen
  \bibfield  {author} {\bibinfo {author} {\bibfnamefont {R.}~\bibnamefont {{Al Maruf}}}, \bibinfo {author} {\bibfnamefont {S.}~\bibnamefont {{Venuturumilli}}}, \bibinfo {author} {\bibfnamefont {D.}~\bibnamefont {{Bharadwaj}}}, \bibinfo {author} {\bibfnamefont {P.}~\bibnamefont {{Anderson}}}, \bibinfo {author} {\bibfnamefont {J.}~\bibnamefont {{Qiu}}}, \bibinfo {author} {\bibfnamefont {Y.}~\bibnamefont {{Yuan}}}, \bibinfo {author} {\bibfnamefont {M.}~\bibnamefont {{Zeeshan}}}, \bibinfo {author} {\bibfnamefont {B.}~\bibnamefont {{Semnani}}}, \bibinfo {author} {\bibfnamefont {P.~J.}\ \bibnamefont {{Poole}}}, \bibinfo {author} {\bibfnamefont {D.}~\bibnamefont {{Dalacu}}}, \bibinfo {author} {\bibfnamefont {K.}~\bibnamefont {{Resch}}}, \bibinfo {author} {\bibfnamefont {M.~E.}\ \bibnamefont {{Reimer}}},\ and\ \bibinfo {author} {\bibfnamefont {M.}~\bibnamefont {{Bajcsy}}},\ }\bibfield  {title} {\bibinfo {title} {{Widely tunable solid-state source of single-photons matching an atomic transition}},\ }\href
  {https://doi.org/10.48550/arXiv.2309.06734} {\bibfield  {journal} {\bibinfo  {journal} {arXiv e-prints}\ ,\ \bibinfo {eid} {arXiv:2309.06734}} (\bibinfo {year} {2023})},\ \Eprint {https://arxiv.org/abs/2309.06734} {arXiv:2309.06734 [quant-ph]} \BibitemShut {NoStop}%
\bibitem [{\citenamefont {Wolters}\ \emph {et~al.}(2017)\citenamefont {Wolters}, \citenamefont {Buser}, \citenamefont {Horsley}, \citenamefont {B\'eguin}, \citenamefont {J\"ockel}, \citenamefont {Jahn}, \citenamefont {Warburton},\ and\ \citenamefont {Treutlein}}]{Wolters:2017}%
  \BibitemOpen
  \bibfield  {author} {\bibinfo {author} {\bibfnamefont {J.}~\bibnamefont {Wolters}}, \bibinfo {author} {\bibfnamefont {G.}~\bibnamefont {Buser}}, \bibinfo {author} {\bibfnamefont {A.}~\bibnamefont {Horsley}}, \bibinfo {author} {\bibfnamefont {L.}~\bibnamefont {B\'eguin}}, \bibinfo {author} {\bibfnamefont {A.}~\bibnamefont {J\"ockel}}, \bibinfo {author} {\bibfnamefont {J.-P.}\ \bibnamefont {Jahn}}, \bibinfo {author} {\bibfnamefont {R.~J.}\ \bibnamefont {Warburton}},\ and\ \bibinfo {author} {\bibfnamefont {P.}~\bibnamefont {Treutlein}},\ }\bibfield  {title} {\bibinfo {title} {{Simple Atomic Quantum Memory Suitable for Semiconductor Quantum Dot Single Photons}},\ }\href {https://doi.org/10.1103/PhysRevLett.119.060502} {\bibfield  {journal} {\bibinfo  {journal} {Phys. Rev. Lett.}\ }\textbf {\bibinfo {volume} {119}},\ \bibinfo {pages} {060502} (\bibinfo {year} {2017})}\BibitemShut {NoStop}%
\bibitem [{\citenamefont {Thomas}\ \emph {et~al.}(2024)\citenamefont {Thomas}, \citenamefont {Wagner}, \citenamefont {Joos}, \citenamefont {Sittig}, \citenamefont {Nawrath}, \citenamefont {Burdekin}, \citenamefont {de~Buy~Wenniger}, \citenamefont {Rasiah}, \citenamefont {Huber-Loyola}, \citenamefont {Sagona-Stophel}, \citenamefont {Höfling}, \citenamefont {Jetter}, \citenamefont {Michler}, \citenamefont {Walmsley}, \citenamefont {Portalupi},\ and\ \citenamefont {Ledingham}}]{Thomas:2024}%
  \BibitemOpen
  \bibfield  {author} {\bibinfo {author} {\bibfnamefont {S.~E.}\ \bibnamefont {Thomas}}, \bibinfo {author} {\bibfnamefont {L.}~\bibnamefont {Wagner}}, \bibinfo {author} {\bibfnamefont {R.}~\bibnamefont {Joos}}, \bibinfo {author} {\bibfnamefont {R.}~\bibnamefont {Sittig}}, \bibinfo {author} {\bibfnamefont {C.}~\bibnamefont {Nawrath}}, \bibinfo {author} {\bibfnamefont {P.}~\bibnamefont {Burdekin}}, \bibinfo {author} {\bibfnamefont {I.~M.}\ \bibnamefont {de~Buy~Wenniger}}, \bibinfo {author} {\bibfnamefont {M.~J.}\ \bibnamefont {Rasiah}}, \bibinfo {author} {\bibfnamefont {T.}~\bibnamefont {Huber-Loyola}}, \bibinfo {author} {\bibfnamefont {S.}~\bibnamefont {Sagona-Stophel}}, \bibinfo {author} {\bibfnamefont {S.}~\bibnamefont {Höfling}}, \bibinfo {author} {\bibfnamefont {M.}~\bibnamefont {Jetter}}, \bibinfo {author} {\bibfnamefont {P.}~\bibnamefont {Michler}}, \bibinfo {author} {\bibfnamefont {I.~A.}\ \bibnamefont {Walmsley}}, \bibinfo {author} {\bibfnamefont {S.~L.}\ \bibnamefont {Portalupi}},\ and\ \bibinfo
  {author} {\bibfnamefont {P.~M.}\ \bibnamefont {Ledingham}},\ }\bibfield  {title} {\bibinfo {title} {Deterministic storage and retrieval of telecom light from a quantum dot single-photon source interfaced with an atomic quantum memory},\ }\href {https://doi.org/10.1126/sciadv.adi7346} {\bibfield  {journal} {\bibinfo  {journal} {Science Advances}\ }\textbf {\bibinfo {volume} {10}},\ \bibinfo {pages} {eadi7346} (\bibinfo {year} {2024})}\BibitemShut {NoStop}%
\bibitem [{\citenamefont {Rastogi}\ \emph {et~al.}(2019)\citenamefont {Rastogi}, \citenamefont {Saglamyurek}, \citenamefont {Hrushevskyi}, \citenamefont {Hubele},\ and\ \citenamefont {LeBlanc}}]{Rastogi:2019}%
  \BibitemOpen
  \bibfield  {author} {\bibinfo {author} {\bibfnamefont {A.}~\bibnamefont {Rastogi}}, \bibinfo {author} {\bibfnamefont {E.}~\bibnamefont {Saglamyurek}}, \bibinfo {author} {\bibfnamefont {T.}~\bibnamefont {Hrushevskyi}}, \bibinfo {author} {\bibfnamefont {S.}~\bibnamefont {Hubele}},\ and\ \bibinfo {author} {\bibfnamefont {L.~J.}\ \bibnamefont {LeBlanc}},\ }\bibfield  {title} {\bibinfo {title} {Discerning quantum memories based on electromagnetically-induced-transparency and autler-townes-splitting protocols},\ }\href {https://doi.org/10.1103/PhysRevA.100.012314} {\bibfield  {journal} {\bibinfo  {journal} {Phys. Rev. A}\ }\textbf {\bibinfo {volume} {100}},\ \bibinfo {pages} {012314} (\bibinfo {year} {2019})}\BibitemShut {NoStop}%
\bibitem [{\citenamefont {Main}\ \emph {et~al.}(2021)\citenamefont {Main}, \citenamefont {Hird}, \citenamefont {Gao}, \citenamefont {Oguz}, \citenamefont {Saunders}, \citenamefont {Walmsley},\ and\ \citenamefont {Ledingham}}]{Main:2021}%
  \BibitemOpen
  \bibfield  {author} {\bibinfo {author} {\bibfnamefont {D.}~\bibnamefont {Main}}, \bibinfo {author} {\bibfnamefont {T.~M.}\ \bibnamefont {Hird}}, \bibinfo {author} {\bibfnamefont {S.}~\bibnamefont {Gao}}, \bibinfo {author} {\bibfnamefont {E.}~\bibnamefont {Oguz}}, \bibinfo {author} {\bibfnamefont {D.~J.}\ \bibnamefont {Saunders}}, \bibinfo {author} {\bibfnamefont {I.~A.}\ \bibnamefont {Walmsley}},\ and\ \bibinfo {author} {\bibfnamefont {P.~M.}\ \bibnamefont {Ledingham}},\ }\bibfield  {title} {\bibinfo {title} {Preparing narrow velocity distributions for quantum memories in room-temperature alkali-metal vapors},\ }\href {https://doi.org/10.1103/PhysRevA.103.043105} {\bibfield  {journal} {\bibinfo  {journal} {Phys. Rev. A}\ }\textbf {\bibinfo {volume} {103}},\ \bibinfo {pages} {043105} (\bibinfo {year} {2021})}\BibitemShut {NoStop}%
\bibitem [{\citenamefont {{D J Croucher}}\ and\ \citenamefont {{J L Clark}}(1969)}]{DJCroucher.1969}%
  \BibitemOpen
  \bibfield  {author} {\bibinfo {author} {\bibnamefont {{D J Croucher}}}\ and\ \bibinfo {author} {\bibnamefont {{J L Clark}}},\ }\bibfield  {title} {\bibinfo {title} {Total collision cross sections and van der waals constants for alkali atom interactions with atoms and non-reactive diatomic molecules at thermal energies},\ }\href {https://doi.org/10.1088/0022-3700/2/5/313} {\bibfield  {journal} {\bibinfo  {journal} {Journal of Physics B: Atomic and Molecular Physics}\ }\textbf {\bibinfo {volume} {2}},\ \bibinfo {pages} {603} (\bibinfo {year} {1969})}\BibitemShut {NoStop}%
\bibitem [{\citenamefont {Esguerra}\ \emph {et~al.}(2023)\citenamefont {Esguerra}, \citenamefont {Me\ss{}ner}, \citenamefont {Robertson}, \citenamefont {Ewald}, \citenamefont {G\"undo\ifmmode~\breve{g}\else \u{g}\fi{}an},\ and\ \citenamefont {Wolters}}]{Esguerra:2023}%
  \BibitemOpen
  \bibfield  {author} {\bibinfo {author} {\bibfnamefont {L.}~\bibnamefont {Esguerra}}, \bibinfo {author} {\bibfnamefont {L.}~\bibnamefont {Me\ss{}ner}}, \bibinfo {author} {\bibfnamefont {E.}~\bibnamefont {Robertson}}, \bibinfo {author} {\bibfnamefont {N.~V.}\ \bibnamefont {Ewald}}, \bibinfo {author} {\bibfnamefont {M.}~\bibnamefont {G\"undo\ifmmode~\breve{g}\else \u{g}\fi{}an}},\ and\ \bibinfo {author} {\bibfnamefont {J.}~\bibnamefont {Wolters}},\ }\bibfield  {title} {\bibinfo {title} {Optimization and readout-noise analysis of a warm-vapor electromagnetically-induced-transparency memory on the cs ${D}_{1}$ line},\ }\href {https://doi.org/10.1103/PhysRevA.107.042607} {\bibfield  {journal} {\bibinfo  {journal} {Phys. Rev. A}\ }\textbf {\bibinfo {volume} {107}},\ \bibinfo {pages} {042607} (\bibinfo {year} {2023})}\BibitemShut {NoStop}%
\bibitem [{\citenamefont {Finkelstein}\ \emph {et~al.}(2021)\citenamefont {Finkelstein}, \citenamefont {Lahad}, \citenamefont {Cohen}, \citenamefont {Davidson}, \citenamefont {Kiriati}, \citenamefont {Poem},\ and\ \citenamefont {Firstenberg}}]{Finkelstein:2021}%
  \BibitemOpen
  \bibfield  {author} {\bibinfo {author} {\bibfnamefont {R.}~\bibnamefont {Finkelstein}}, \bibinfo {author} {\bibfnamefont {O.}~\bibnamefont {Lahad}}, \bibinfo {author} {\bibfnamefont {I.}~\bibnamefont {Cohen}}, \bibinfo {author} {\bibfnamefont {O.}~\bibnamefont {Davidson}}, \bibinfo {author} {\bibfnamefont {S.}~\bibnamefont {Kiriati}}, \bibinfo {author} {\bibfnamefont {E.}~\bibnamefont {Poem}},\ and\ \bibinfo {author} {\bibfnamefont {O.}~\bibnamefont {Firstenberg}},\ }\bibfield  {title} {\bibinfo {title} {Continuous protection of a collective state from inhomogeneous dephasing},\ }\href {https://doi.org/10.1103/PhysRevX.11.011008} {\bibfield  {journal} {\bibinfo  {journal} {Phys. Rev. X}\ }\textbf {\bibinfo {volume} {11}},\ \bibinfo {pages} {011008} (\bibinfo {year} {2021})}\BibitemShut {NoStop}%
\bibitem [{\citenamefont {Gorshkov}\ \emph {et~al.}(2007)\citenamefont {Gorshkov}, \citenamefont {Andr\'e}, \citenamefont {Lukin},\ and\ \citenamefont {S\o{}rensen}}]{GorshkovPhysRevA.76.033806}%
  \BibitemOpen
  \bibfield  {author} {\bibinfo {author} {\bibfnamefont {A.~V.}\ \bibnamefont {Gorshkov}}, \bibinfo {author} {\bibfnamefont {A.}~\bibnamefont {Andr\'e}}, \bibinfo {author} {\bibfnamefont {M.~D.}\ \bibnamefont {Lukin}},\ and\ \bibinfo {author} {\bibfnamefont {A.~S.}\ \bibnamefont {S\o{}rensen}},\ }\bibfield  {title} {\bibinfo {title} {Photon storage in $\ensuremath{\Lambda}$-type optically dense atomic media. iii. effects of inhomogeneous broadening},\ }\href {https://doi.org/10.1103/PhysRevA.76.033806} {\bibfield  {journal} {\bibinfo  {journal} {Phys. Rev. A}\ }\textbf {\bibinfo {volume} {76}},\ \bibinfo {pages} {033806} (\bibinfo {year} {2007})}\BibitemShut {NoStop}%
\bibitem [{\citenamefont {{\DJ}uji{\'c}}\ \emph {et~al.}(2024)\citenamefont {{\DJ}uji{\'c}}, \citenamefont {Buhin}, \citenamefont {{\v{S}}anti{\'c}}, \citenamefont {Aumiler},\ and\ \citenamefont {Ban}}]{Dujic.2024}%
  \BibitemOpen
  \bibfield  {author} {\bibinfo {author} {\bibfnamefont {M.}~\bibnamefont {{\DJ}uji{\'c}}}, \bibinfo {author} {\bibfnamefont {D.}~\bibnamefont {Buhin}}, \bibinfo {author} {\bibfnamefont {N.}~\bibnamefont {{\v{S}}anti{\'c}}}, \bibinfo {author} {\bibfnamefont {D.}~\bibnamefont {Aumiler}},\ and\ \bibinfo {author} {\bibfnamefont {T.}~\bibnamefont {Ban}},\ }\bibfield  {title} {\bibinfo {title} {Comparative analysis of light storage in antirelaxation-coated and buffer-gas-filled alkali vapor cells},\ }\href {https://doi.org/10.1038/s41598-024-63489-8} {\bibfield  {journal} {\bibinfo  {journal} {Scientific Reports}\ }\textbf {\bibinfo {volume} {14}},\ \bibinfo {pages} {14467} (\bibinfo {year} {2024})}\BibitemShut {NoStop}%
\bibitem [{\citenamefont {Phillips}\ \emph {et~al.}(2008)\citenamefont {Phillips}, \citenamefont {Gorshkov},\ and\ \citenamefont {Novikova}}]{Phillips:2008}%
  \BibitemOpen
  \bibfield  {author} {\bibinfo {author} {\bibfnamefont {N.~B.}\ \bibnamefont {Phillips}}, \bibinfo {author} {\bibfnamefont {A.~V.}\ \bibnamefont {Gorshkov}},\ and\ \bibinfo {author} {\bibfnamefont {I.}~\bibnamefont {Novikova}},\ }\bibfield  {title} {\bibinfo {title} {Optimal light storage in atomic vapor},\ }\href {https://doi.org/10.1103/PhysRevA.78.023801} {\bibfield  {journal} {\bibinfo  {journal} {Phys. Rev. A}\ }\textbf {\bibinfo {volume} {78}},\ \bibinfo {pages} {023801} (\bibinfo {year} {2008})}\BibitemShut {NoStop}%
\bibitem [{\citenamefont {Thoma}\ \emph {et~al.}(2016)\citenamefont {Thoma}, \citenamefont {Schnauber}, \citenamefont {Gschrey}, \citenamefont {Seifried}, \citenamefont {Wolters}, \citenamefont {Schulze}, \citenamefont {Strittmatter}, \citenamefont {Rodt}, \citenamefont {Carmele}, \citenamefont {Knorr}, \citenamefont {Heindel},\ and\ \citenamefont {Reitzenstein}}]{Thoma:2016}%
  \BibitemOpen
  \bibfield  {author} {\bibinfo {author} {\bibfnamefont {A.}~\bibnamefont {Thoma}}, \bibinfo {author} {\bibfnamefont {P.}~\bibnamefont {Schnauber}}, \bibinfo {author} {\bibfnamefont {M.}~\bibnamefont {Gschrey}}, \bibinfo {author} {\bibfnamefont {M.}~\bibnamefont {Seifried}}, \bibinfo {author} {\bibfnamefont {J.}~\bibnamefont {Wolters}}, \bibinfo {author} {\bibfnamefont {J.-H.}\ \bibnamefont {Schulze}}, \bibinfo {author} {\bibfnamefont {A.}~\bibnamefont {Strittmatter}}, \bibinfo {author} {\bibfnamefont {S.}~\bibnamefont {Rodt}}, \bibinfo {author} {\bibfnamefont {A.}~\bibnamefont {Carmele}}, \bibinfo {author} {\bibfnamefont {A.}~\bibnamefont {Knorr}}, \bibinfo {author} {\bibfnamefont {T.}~\bibnamefont {Heindel}},\ and\ \bibinfo {author} {\bibfnamefont {S.}~\bibnamefont {Reitzenstein}},\ }\bibfield  {title} {\bibinfo {title} {Exploring dephasing of a solid-state quantum emitter via time- and temperature-dependent {H}ong-{O}u-{M}andel experiments},\ }\href {https://doi.org/10.1103/PhysRevLett.116.033601}
  {\bibfield  {journal} {\bibinfo  {journal} {Phys. Rev. Lett.}\ }\textbf {\bibinfo {volume} {116}},\ \bibinfo {pages} {033601} (\bibinfo {year} {2016})}\BibitemShut {NoStop}%
\bibitem [{\citenamefont {Gao}\ \emph {et~al.}(2019)\citenamefont {Gao}, \citenamefont {Lazo-Arjona}, \citenamefont {Brecht}, \citenamefont {Kaczmarek}, \citenamefont {Thomas}, \citenamefont {Nunn}, \citenamefont {Ledingham}, \citenamefont {Saunders},\ and\ \citenamefont {Walmsley}}]{Gao:2019}%
  \BibitemOpen
  \bibfield  {author} {\bibinfo {author} {\bibfnamefont {S.}~\bibnamefont {Gao}}, \bibinfo {author} {\bibfnamefont {O.}~\bibnamefont {Lazo-Arjona}}, \bibinfo {author} {\bibfnamefont {B.}~\bibnamefont {Brecht}}, \bibinfo {author} {\bibfnamefont {K.~T.}\ \bibnamefont {Kaczmarek}}, \bibinfo {author} {\bibfnamefont {S.~E.}\ \bibnamefont {Thomas}}, \bibinfo {author} {\bibfnamefont {J.}~\bibnamefont {Nunn}}, \bibinfo {author} {\bibfnamefont {P.~M.}\ \bibnamefont {Ledingham}}, \bibinfo {author} {\bibfnamefont {D.~J.}\ \bibnamefont {Saunders}},\ and\ \bibinfo {author} {\bibfnamefont {I.~A.}\ \bibnamefont {Walmsley}},\ }\bibfield  {title} {\bibinfo {title} {Optimal coherent filtering for single noisy photons},\ }\href {https://doi.org/10.1103/PhysRevLett.123.213604} {\bibfield  {journal} {\bibinfo  {journal} {Phys. Rev. Lett.}\ }\textbf {\bibinfo {volume} {123}},\ \bibinfo {pages} {213604} (\bibinfo {year} {2019})}\BibitemShut {NoStop}%
\end{thebibliography}%

\end{document}